\documentclass[11pt]{article}

\usepackage[a4paper,margin=2.4cm]{geometry}
\usepackage{newtxtext}
\usepackage{indentfirst}
\usepackage{microtype}
\usepackage{xcolor}
\usepackage{graphicx}
\usepackage{float}
\usepackage[font=small,labelfont=bf,labelsep=period]{caption}
\usepackage{booktabs}
\usepackage{tabularx}
\usepackage{array}
\usepackage{calc}
\usepackage{multirow}
\usepackage{changepage}
\usepackage{pdflscape}
\usepackage{amsmath,amssymb,amsthm}
\usepackage{newtxmath}
\usepackage[natbibapa]{apacite}
\usepackage{url}
\usepackage{titlesec}
\usepackage[hidelinks]{hyperref}

\graphicspath{{figures/}}
\titlespacing*{\section}{0pt}{2.5ex plus .6ex minus .2ex}{1.0ex}
\titlespacing*{\subsection}{0pt}{2.0ex plus .5ex minus .2ex}{0.7ex}
\newlength{\fulllength}
\newlength{\extralength}
\newlength{\cellWidtha}
\newlength{\cellWidthb}
\newlength{\cellWidthc}
\newlength{\cellWidthd}
\newlength{\cellWidthe}
\newlength{\cellWidthf}
\newlength{\cellWidthg}
\newlength{\cellWidthh}
\newcolumntype{C}{>{\centering\arraybackslash}X}
\newtheorem{Hypothesis}{Hypothesis}

\title{Measuring Psychological Constructs from Social Media Text Using the Word Embedding Projection Approach}
\author{Xudong Deng\thanks{ORCID: \href{https://orcid.org/0000-0001-7391-8865}{0000-0001-7391-8865}. Corresponding author: \href{mailto:18b910029@stu.hit.edu.cn}{18b910029@stu.hit.edu.cn}.}
\and Yijun Li\thanks{Email: \href{mailto:liyijun@hit.edu.cn}{liyijun@hit.edu.cn}.}\\[0.5em]
\small School of Management, Harbin Institute of Technology, Harbin 150001, China}
\date{}

\begin{document}
\maketitle

\begin{center}
\small
Published in \textit{Behavioral Sciences} (2026),
\textbf{16}(5), 762.\\
\href{https://doi.org/10.3390/bs16050762}
{https://doi.org/10.3390/bs16050762}
\end{center}

\begin{abstract}
Social media text offers new opportunities for tracking psychological constructs over time, but such measurement requires methods that are interpretable and stable in longitudinal settings. Grounded in Conceptual Space Theory and semantic projection, this study develops the \emph{Word Embedding Projection Approach} (WEPA), which represents psychological constructs as semantic axes defined by theory-driven anchor words in a domain-specific word embedding space. WEPA scores are interpreted as indicators of expressed psychological salience in user-generated text. Using panel data from Keep, a Chinese fitness-oriented social media platform (177,829 users; 2,668,298 weekly observations), we evaluate WEPA across four studies. Study I examines the measurement foundation through anchor-word validation, human-coded benchmarks, and cross-temporal semantic-axis stability checks. Studies II and III show theoretically interpretable predictive associations between WEPA-derived scores and subsequent exercise duration for goal-setting constructs and the four source dimensions of self-efficacy. Study IV provides exploratory relative-week trajectories of aggregate construct dynamics. Overall, WEPA shows strong agreement with human-coded benchmarks for goal specificity and physiological states and achieves broader text coverage than a dictionary-based baseline. These findings suggest that WEPA offers a promising approach to theory-driven psychological measurement from domain-specific social media text.
\end{abstract}


\section{Introduction}
\label{sec:introduction}

User-generated content on social media offers new opportunities for large-scale, non-intrusive psychological measurement because it contains rich digital traces of users' thoughts, feelings, and behavior \citep{golder2014DigitalFootprints, Eichstaedt2020Tracking, berger2020UnitingTribes, Feng2023SNS, sultan2023leaving, Zhang2024PsyOwnershipUGC}. Word embedding methods have been used to study macro-level phenomena such as stereotypes, ideology, and cultural change \citep{kozlowski2019GeometryCulture, grand2022SemanticProjection}. However, their validity for measuring individual psychological constructs, such as self-efficacy and goal commitment, remains less clear, especially in longitudinal settings where scores must remain interpretable and comparable over time. Dictionary-based methods, including LIWC-style tools, provide transparent lexical indicators but rely on closed vocabularies and exact word matching, which can lead to sparse coverage in short texts \citep{Hartmann2019Comparing, Eichstaedt2020ClosedOpen}. Contextual language models and large language models offer richer semantic representations, but their outputs can be difficult to interpret as fixed construct dimensions and may be sensitive to context, prompts, model updates, or representational drift \citep{Demszky2023Using, Sclar2023Quantifying, Zhou2024Larger, nikadon2025BERTAgent-MeasurePsyMetrics}.

{To address this measurement problem, we extend semantic projection to individual-level psychological construct measurement and develop the \emph{Word Embedding Projection Approach} (WEPA). WEPA does not introduce semantic projection as a new computational principle; instead, it adapts the geometric logic of word embeddings to a psychometric measurement task.} In WEPA, theory-driven anchor words define semantic axes, and user text is projected onto these axes to obtain construct scores. This design links Conceptual Space Theory with domain-specific static embeddings, localized anchor-word dictionaries, and validation procedures to form an interpretable workflow for measuring psychological constructs from platform text. {In this study, WEPA scores are interpreted as indicators of expressed psychological salience in user-generated text. They capture how strongly users' language reflects construct-related meanings, not direct observations of latent psychological states or external motivational conditions}.

WEPA makes three specific contributions. First, it applies semantic projection to individual-level psychological measurement in a large-scale Chinese digital fitness community, where short texts and domain-specific expressions challenge closed-vocabulary methods. Second, it demonstrates context-specific anchor-word development and validation, using Goal-Setting Theory and Social Cognitive Theory to guide the operationalization of psychological constructs in Chinese fitness-community text. {Third, it treats semantic projection as a measurement workflow, not only a scoring technique, by examining measurement foundation, criterion validity, and cross-temporal semantic-axis stability. In this sense, WEPA should be understood as a psychometric implementation of semantic projection, not as a universal replacement for dictionary-based, supervised-learning or contextual language-model approaches.}

We evaluate WEPA using an observational design and a stepwise validation strategy. Our analysis draws on panel data from Keep, a Chinese digital fitness community, comprising 177,829 users and 2,668,298 weekly observations. Study I establishes the measurement foundation through anchor-word validation, human-coded benchmarks, and cross-temporal semantic-axis stability checks. {Studies II and III} examine criterion validity by testing whether WEPA-derived construct scores are associated with subsequent exercise duration for goal-setting constructs and the four source dimensions of self-efficacy. \mbox{Study IV} provides an exploratory relative-week analysis of aggregate construct trajectories after aligning users by weeks since registration. {Because the study is observational and based on a single platform, the findings should be interpreted as predictive associations and exploratory descriptive patterns, not as causal effects or universal psychological dynamics.} To support transparent and reproducible research, we provide an open-source implementation of the WEPA scoring procedure.

\section{Theoretical Foundations and WEPA Design}
\label{sec:theory_algorithm}

\subsection{Text as a Basis for Psychological Measurement}
\label{subsec:measurement_foundation}
Language is central to human social interaction and systematically reflects personality traits, behavioral intentions, and psychological tendencies \citep{pennebaker2015development, herderich2024computational}. Research at the intersection of linguistics and psychology has shown that patterns of language use are associated with cognitive structure, affective states, motivational orientation, and related psychological processes \citep{Humphreys2017Automated, Kjell2019Semantic, Lee2023LDA, Luisa2024PsyChange}. This body of work provides a foundation for using social media text to measure individual psychological constructs.

In psychometrics, psychological constructs are latent variables that cannot be observed directly and must be inferred from observable indicators. In classical reflective models, indicators are assumed to be caused by the latent construct and are therefore expected to be correlated and partly interchangeable because they share the same underlying source. Under this logic, adding or removing a small number of indicators affects measurement precision, not the ontological meaning of the construct itself \citep{Kyle2020Tools, Flake2017ConstructValidation}. By extension, digital traces embedded in user-generated content can be treated as reflective linguistic indicators. Lexical choices and broader patterns of language use can thus be understood as manifestations of latent constructs such as self-efficacy and goal commitment, not as constitutive components of those constructs.

Treating user-generated content as reflective indicators implies three assumptions. First, latent psychological constructs give rise to recurring linguistic features and behavioral expressions. In a fitness context, for example, individuals with high self-efficacy may be more likely to use words such as \textit{capable} and \textit{confident}. Second, different textual features, including lexical choice and semantic orientation, can serve as related indicators because they share common variance generated by the same latent construct. Third, adding or removing individual words or expressions changes measurement reliability rather than the theoretical boundaries of the construct. Prior research has used this logic to identify latent structures linking language use to personality and emotion, suggesting that language-based indicators can capture psychological patterns with temporal regularity and behavioral relevance \citep{golder2014DigitalFootprints, Eichstaedt2020Tracking}.

\subsection{Conceptual Space and Construct Representation}
\label{subsec:concept_space}
\begin{figure}[!htbp]
    \begin{adjustwidth}{-\extralength}{0cm}
        \centering
        \includegraphics[width=\fulllength]{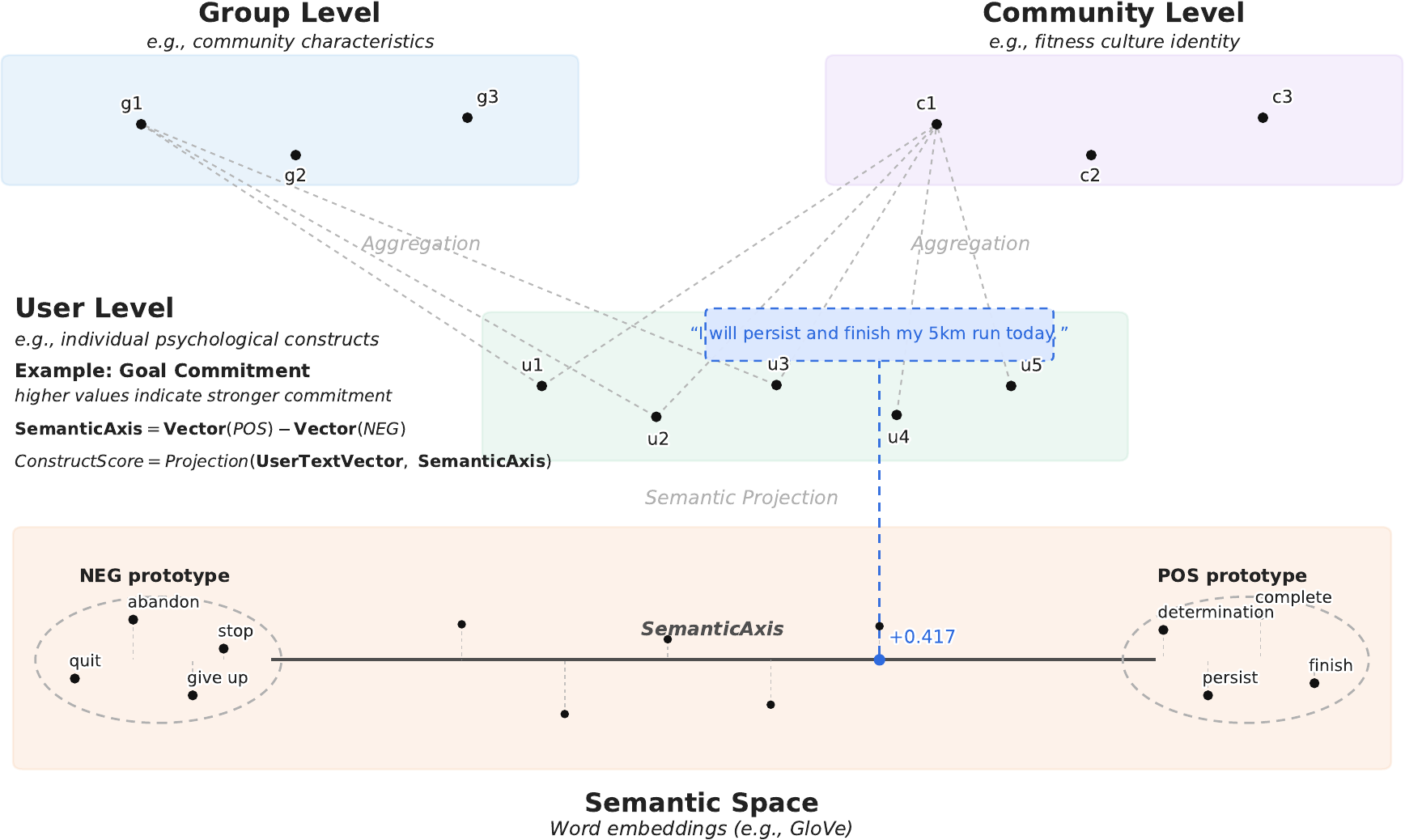}
    \end{adjustwidth}
    \caption{{Conceptual-space 
 representation of text-based psychological measurement. Word embeddings define the semantic space, theory-driven anchor words define bipolar construct axes, and user text is projected onto these axes to generate observable construct indicators. Aggregated scores can then be used to describe group- or community-level patterns.}}
    \label{fig:chp3-concept-space-hierarchy}
\end{figure}

Conceptual Space Theory treats abstract concepts as mental representations that can be formalized in a multidimensional geometric space \citep{aceves2024mobilizing, Douven2011Vagueness}. In WEPA, word embeddings define this semantic space at the word level, and theory-driven anchor words define construct-specific semantic axes. User text is then projected onto these axes to generate construct scores, which can later be aggregated to describe group- or community-level patterns when appropriate, as illustrated in Figure~\ref{fig:chp3-concept-space-hierarchy}.

{The core operation of WEPA is the mapping from word-level semantic representations to user-level construct indicators. In psychometric terms, constructs are the latent psychological variables of interest, dimensions are their theoretical components, and indicators are the projection scores generated from text. Conceptual Space Theory formalizes the semantic structure of a construct, while psychometric theory clarifies that the latent construct itself is inferred from observable indicators. Thus, a bipolar semantic axis does not replace the latent construct. It defines a theory-guided measurement direction in language space, and the projection score serves as a text-based indicator of expressed construct salience.}

{A \emph{bipolar dimension} refers to a structure in which opposing tendencies on a single latent construct, such as \emph{extraversion--introversion} or \emph{high pleasure--low pleasure}, are treated as opposite ends of the same continuum \citep{Barchard2024Testing, Reise2013Scoring}.} Goal commitment illustrates this structure. High commitment may be represented by prototypes such as \emph{persist}, \emph{determination}, and \emph{complete}, while low commitment may be represented by \emph{quit}, \emph{abandon}, and \emph{give up}. These semantically opposed prototypes define the geometric axis of the construct, and a user's text can be projected onto the axis to obtain a reflective linguistic indicator. For example, the expression, "\emph{I will persist and finish my 5K run today}" falls on the positive side of the axis, indicating relatively high expressed goal commitment.

When a construct contains multiple theoretically distinct dimensions, these dimensions can be represented separately and, when justified, summarized into an overall score \citep{Reise2013Scoring, Rodriguez2016Applying}. Self-efficacy illustrates this case. In Social Cognitive Theory, self-efficacy is shaped by mastery experience, vicarious experience, social persuasion, and physiological and affective states \citep{Bandura1977SelfEfficacy}. In this study, these four sources are operationalized as text-based dimensions that can be examined separately and aggregated as an overall self-efficacy indicator, allowing WEPA to quantify multidimensional constructs while preserving their internal organization.

\subsection{{Measurement Stability and Longitudinal Comparability}}
\label{subsec:measurement_invariance}
{Measurement stability and longitudinal comparability are central concerns when psychological constructs are measured repeatedly over time. A related but stricter concept is measurement invariance, which refers to whether a measurement approach preserves a consistent construct scale across time, groups, or contexts \citep{Liu2017Testing, Leitgob2022Measurement}. In longitudinal text analysis, observed changes are interpretable only when they reflect changes in construct-related expression, not shifts in word meaning or instability in the measurement procedure.}

{Strict measurement invariance is difficult to establish in text-based settings because language evolves and contextualized models may generate representations that vary with input context. WEPA therefore adopts an operational strategy focused on stability and comparability. It uses static word embeddings, fixed semantic axes, and anchor-word aggregation.} Static embeddings provide fixed word vectors after training. Fixed axes keep the measurement direction constant across samples and time. Anchor-word aggregation reduces dependence on any single period-specific expression by representing each pole as a prototype rather than as an individual word.

{These design choices are intended to support stable measurement, but they do not constitute a full test of measurement invariance in the strict psychometric sense. We therefore examine cross-temporal semantic-axis stability in Section~\ref{subsubsec:cross-temporal-stability-of-axis}, focusing on both axis orientation and the rank-order consistency of anchor words over time.}

\subsection{{Alternative Text-Based Measurement Approaches}}
\label{subsec:alternative-text-based-measurement}
Text-based psychological measurement can be implemented through several methodological families. {Dictionary-based approaches, such as LIWC-style lexicons, are transparent and easy to apply, but they depend on predefined word lists and may have limited coverage in domain-specific social media text} \citep{pennebaker2015development, Hartmann2019Comparing, Eichstaedt2020ClosedOpen}. Supervised machine-learning models can improve task-specific prediction when large labeled datasets are available, although they require construct-specific annotations and often provide limited interpretability as theory-defined construct scores \citep{Kadhim2019Survey, Speer2024Taking}. Transformer-based and large language models offer strong semantic flexibility, but their outputs may be sensitive to prompts, model versions, proprietary training procedures, and computational cost \citep{Demszky2023Using, nikadon2025BERTAgent-MeasurePsyMetrics, Sclar2023Quantifying, Zhou2024Larger}. Semantic projection methods, including WEPA, occupy an intermediate position by representing meaning continuously in embedding space while maintaining an explicit link between construct definitions and measurement rules \citep{kozlowski2019GeometryCulture, grand2022SemanticProjection}.

WEPA builds on the semantic-projection tradition and adapts it to theory-driven psychological measurement. It is designed for constructs that can be represented through theoretically meaningful semantic poles and extends anchor-word-based measurement by projecting text in embedding space instead of relying on exact word matching. {For this reason, the dictionary-based method provides the closest baseline for the benchmark analysis in Study I. Broader comparisons with supervised classifiers, transformer encoders, and LLM-based scoring remain valuable directions for future research when high-quality labeled data and stable model access are available. Table~\ref{tab:alternative_text_measurement} summarizes the main strengths and limitations of these approaches and positions WEPA within the broader landscape of text-based psychological measurement.}

\begin{table}[!htb]\footnotesize
\caption{{Alternative approaches to text-based psychological measurement.}}
\label{tab:alternative_text_measurement}
\begin{adjustwidth}{-\extralength}{0cm}
\setlength{\cellWidtha}{\fulllength/3-2\tabcolsep-0in}
\setlength{\cellWidthb}{\fulllength/3-2\tabcolsep-0in}
\setlength{\cellWidthc}{\fulllength/3-2\tabcolsep-0in}

\scalebox{1}[1]{\begin{tabularx}{\fulllength}{>{\raggedright\arraybackslash}m{\cellWidtha}>{\raggedright\arraybackslash}m{\cellWidthb}>{\raggedright\arraybackslash}m{\cellWidthc}}
\toprule
\textbf{{Approach}} & \textbf{{Main Strengths}} & \textbf{{Main Limitations}} \\
\midrule

{Dictionary-based methods}
& {Transparent, interpretable, and computationally efficient}
& {Limited by predefined categories, exact word matching, and domain-specific coverage} \\

{Supervised machine learning}
& {Strong task-specific prediction when high-quality labeled data are available}
& {Requires construct-specific labels; model outputs may be less interpretable as theory-defined construct scores} \\

{Transformer or LLM-based scoring}
& {Flexible semantic understanding and strong contextual language processing}
& {Sensitive to prompts, model versions, computational cost, and reproducibility constraints} \\

{Semantic projection methods, including WEPA}
& {Continuous semantic scoring with explicit anchors and interpretable geometric structure}
& {Require meaningful semantic axes, validated anchors, and stability diagnostics} \\

\bottomrule
\end{tabularx}}
\end{adjustwidth}
\noindent{\footnotesize{{\textit{{Note
:}} {This table summarizes conceptual trade-offs among method families and does not report supervised benchmarks across all model families.}}}}

\end{table}

\subsection{Computational Implementation of WEPA}
\label{subsec:wepa_algorithm}
Having clarified the measurement rationale and the trade-offs among alternative approaches, we now describe how WEPA constructs semantic axes, vectorizes user text, and computes projection-based construct scores.

\subsubsection{Core Computational Procedure}
\label{subsubsec:core_computation}

WEPA operationalizes semantic projection in four steps. First, the positive and negative prototype vectors are computed by averaging the word vectors in the corresponding anchor-word sets: 
\begin{equation}
\mathbf{proto\_vec}_d = \frac{1}{|A_d|} \sum_{a \in A_d} \operatorname{Embed}(a), \quad d \in \{pos, neg\}
\label{eq:proto_vec}
\end{equation}

Second, the difference between the two prototype vectors is normalized to obtain the unit semantic-axis vector:
\begin{equation}
\mathbf{axis\_vec} = \frac{\mathbf{proto\_vec}_{pos} - \mathbf{proto\_vec}_{neg}}{\|\mathbf{proto\_vec}_{pos} - \mathbf{proto\_vec}_{neg}\|}
\label{eq:axis_vec}
\end{equation}

Third, the mean of all valid word vectors in the user's text is computed:
\begin{equation}
\mathbf{text\_vec} = \frac{1}{|W|} \sum_{w \in W} \operatorname{Embed}(w)
\label{eq:text_vec}
\end{equation}

Finally, projecting the text vector onto the semantic axis yields the WEPA score:
\begin{equation}
\text{wepa\_score} = \mathbf{text\_vec} \cdot \mathbf{axis\_vec}
\label{eq:wepa_score}
\end{equation}

Here, $A_{pos}$ and $A_{neg}$ denote the positive and negative anchor-word sets, $\mathbf{proto_vec}_d$ denotes the centroid for polarity $d$, and $W$ contains the cleaned tokens after preprocessing and stop-word removal. The embedding function $\operatorname{Embed}(\cdot)$ returns the learned vector for in-vocabulary tokens and a zero vector for out-of-vocabulary (OOV) tokens. OOV tokens remain in the denominator but add no directional semantic information, pulling low-coverage texts closer to the neutral point. We examine the implications of this choice in the textual-sparsity sensitivity analysis, as discussed in Section~\ref{subsec:sensitivity-to-textual-sparsity}.

WEPA uses a single bipolar semantic axis for each construct. Positive values indicate stronger alignment with the positive pole, and larger absolute values indicate greater deviation from the neutral point. Substantively, these scores capture construct-related linguistic salience. A high score means that a user's language is positioned closer to one semantic pole than the other. This distinction matters when classical theories refer to external conditions, such as challenging goals or social encouragement, while social media text records how users express these experiences. In the remainder of the paper, we refer to these WEPA-derived scores as construct scores unless clarification is needed.

\subsubsection{Implementation Steps}
\label{subsubsec:implementation}

Table~\ref{tab:chap3-wepa-steps} summarizes the WEPA workflow used in this study.

For corpus preparation and embedding training, we build a cleaned 816\,MB corpus from fitness-community text. Chinese text is preprocessed by removing URLs, @-mentions, nonlinguistic symbols, and one-character tokens. We then train domain-specific static word vectors with \texttt{cntext} (v2.2.0) \citep{deng2022cntext}. Based on pilot comparisons, the final GloVe model uses 200 dimensions, a context window of 15 words, and 20 training iterations. Embedding quality is assessed through synonym retrieval for 50 high-frequency fitness terms, with 90\% of retrieved neighbors judged semantically relevant in the fitness domain.

{The WEPA scoring procedure is implemented in \texttt{cntext}, an open-source Python text-analysis package developed and maintained by the first author. The package provides functions for preprocessing, embedding training, semantic-axis construction, and text projection. To support reproducibility, the repository includes installation instructions, a minimal WEPA example, API descriptions, and basic tests for semantic-axis construction and text scoring.} When applying WEPA to new constructs or domains, researchers should focus especially on theoretically grounded construct definitions and rigorous anchor-word development and validation.

\begin{table}[!htb]
\small
\caption{WEPA implementation workflow.}
\label{tab:chap3-wepa-steps}
\begin{adjustwidth}{-\extralength}{0cm}
\renewcommand{\arraystretch}{1.15}
\begin{tabularx}{\fulllength}{
>{\centering\arraybackslash}p{0.5cm}
>{\raggedright\arraybackslash}p{3.5cm}
X}
\toprule
\textbf{Step} & \textbf{Stage} & \textbf{Description} \\
\midrule

1 & Construct definition 
& Specify the theoretical construct and its dimensional structure. The construct should have a clear theoretical definition and, where appropriate, a bipolar interpretation. \\

2 & Anchor construction 
& Select and review anchor words for the construct poles. Anchor words should be theory-driven, domain-sensitive, and expert-reviewed to support content validity. \\

3 & Corpus preparation 
& Collect and clean the domain corpus. The corpus should represent the target language environment and be denoised before embedding training. \\

4 & Embedding training
& Train a domain-specific static word embedding model and assess whether the resulting semantic space captures domain-relevant relations through semantic-neighbor inspection. \\

5 & Semantic projection 
& Construct semantic axes from anchor centroids, compute mean vector of cleaned text tokens, and project text vector onto the axis to obtain WEPA construct score. \\

6 & Measurement validation 
& Evaluate whether scores show theoretically expected measurement properties through anchor validation, benchmark comparison, temporal stability checks, and criterion-validity analyses. \\

7 & Empirical application 
& Use the validated scores in substantive analyses, such as panel regressions, predictive association models, or exploratory longitudinal analyses. \\

\bottomrule
\end{tabularx}
\end{adjustwidth}
\end{table}

\section{Validation Strategy, Data, and Measurement}
\label{sec:verification-framework-study-design}

\subsection{Validation Strategy}
\label{subsec:validation_strategy}
\begin{figure}[!htbp]
\begin{adjustwidth}{-\extralength}{0cm}
\centering
\includegraphics[width=\fulllength]{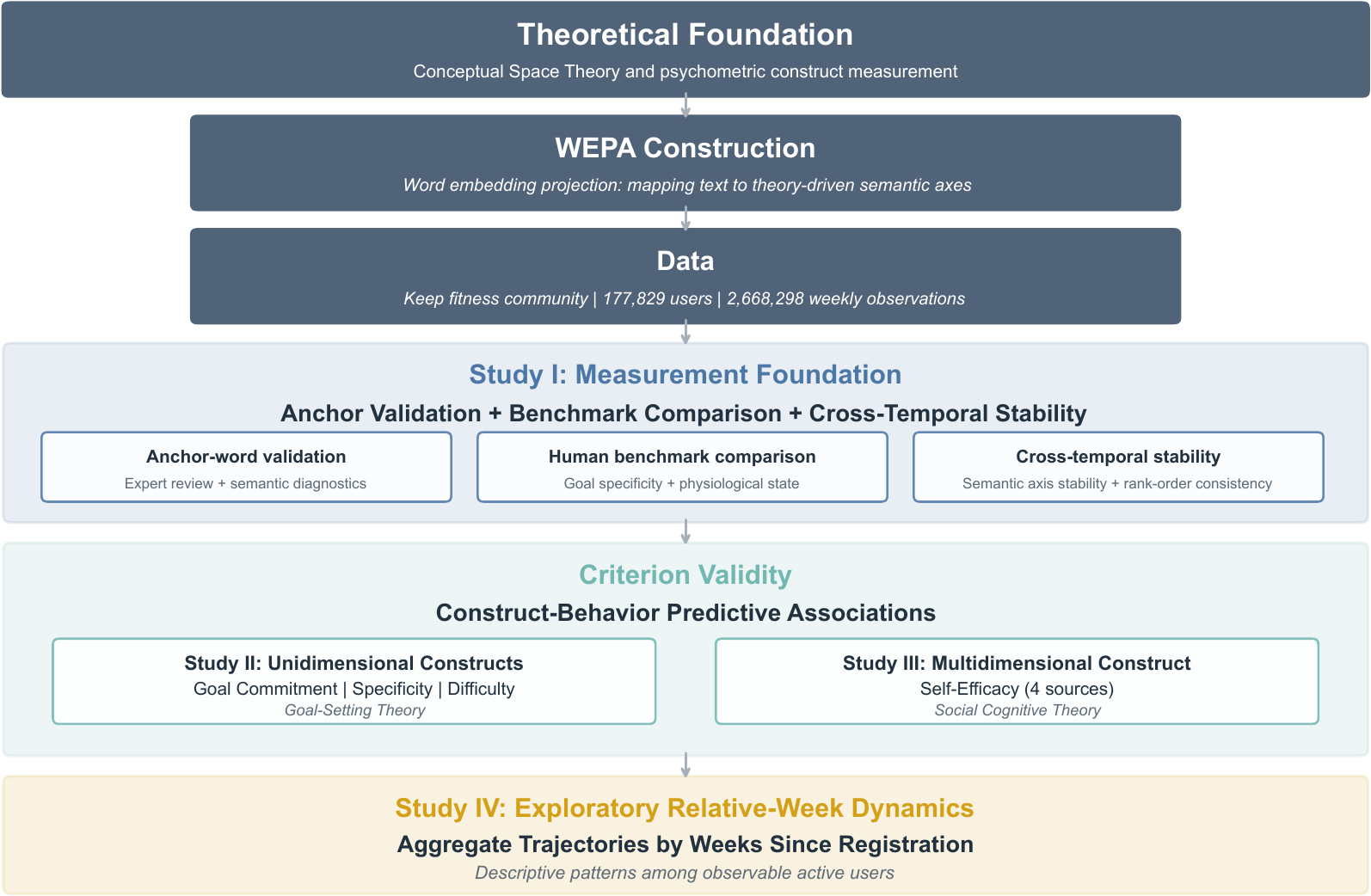}
\end{adjustwidth}
\caption{Three-layer 
validation framework.}
\label{fig:research-framework}
\end{figure}

WEPA is evaluated through a structured validation framework that combines measurement validation with exploratory longitudinal analysis (see Figure~\ref{fig:research-framework}). Study I establishes the measurement foundation through anchor-word validation, benchmark-based evaluation against human annotations, and cross-temporal stability assessment. Studies II and III examine criterion validity by testing whether construct scores are associated with subsequent exercise behavior. Study IV provides an exploratory descriptive analysis of relative-week dynamics by aligning observed users according to weeks since registration and examining aggregate construct trajectories over time. Across the four studies, the analysis proceeds from measurement foundation to criterion validity and then to exploratory evidence on longitudinal psychological dynamics in a naturalistic digital setting.

\subsection{Data and Sample}
\label{subsec:data_sample}

The data come from Keep, a leading fitness app in China that combines exercise tracking with social functions such as posts, check-ins, and topic discussions. This setting provides both user-generated text containing psychological cues and behavioral records, such as exercise duration, that can serve as criterion measures.

We collected publicly accessible secondary digital trace data from 1 February 2015 to \mbox{1 June 2021}. {The sample was expanded from active seed users through follower and followee networks, with duplicate records removed during data collection. This network-based procedure was intended to broaden coverage across active-user neighborhoods, not to produce a probability sample of the entire platform population. Data collection involved no direct interaction with users and accessed no private or restricted content.}

We cleaned the raw data and aggregated them to the user--week level. Weekly aggregation increases text density, reduces day-level noise, and aligns with the weekly rhythm of exercise behavior. Exercise duration and other numerical indicators were summed by week, while all posts published by a user within a given week were concatenated chronologically to form the weekly text. {Before analysis, user identifiers were irreversibly hashed, personally identifiable information was removed, and results were reported only in aggregate form.} We removed user--week observations with empty text and excluded users with missing registration information or fewer than two valid observation weeks, yielding a final sample of 177,829 users and 2,668,298 weekly observations.

\subsection{Variable Measurement}
\label{subsec:variable_measurement}
Exercise engagement intensity is measured by weekly cumulative exercise duration (\textit{ExDur}, minutes per week), transformed as $\ln(x+1)$ to reduce right skew. This variable captures behavioral investment and serves as the objective behavioral criterion in the validation analyses.

The independent variables are WEPA-based construct scores derived from users' weekly text. The raw WEPA score is the scalar projection of the user--week text vector onto a construct-specific semantic axis, as defined in Equation~(\ref{eq:wepa_score}). The goal-setting constructs include goal commitment, goal specificity, and perceived goal difficulty. Self-efficacy is measured using Bandura's four-source model, including mastery experience, vicarious experience, social persuasion, and physiological states. The sum of these four dimensions forms the overall self-efficacy score, while the dimensions are also examined separately.

All control variables are time-varying and cover demographic characteristics, expressive volume, behavioral inertia, and social incentives. User age (\textit{age
}) is measured in years as the difference between the observation week and the user's date of birth. Expressive volume (\textit{StrLen}) is the number of characters posted during the week. Behavioral inertia (\textit{DurChg}) is the first difference in logged exercise duration between adjacent weeks. Social engagement (\textit{SocEngage}) is the weekly sum of discussion and check-in activities, and social feedback (\textit{SocFeedback}) is the weekly sum of likes and comments received. Both social variables are transformed as $\ln(x+1)$. All continuous predictors, including WEPA scores and controls, are standardized in regression analyses unless otherwise stated.

For panel identification, \textit{userid} is used as the anonymized user identifier, and \textit{WeekI} records the number of weeks between the current observation week and the user's registration week. Tables~\ref{tab:chap3-dataset-fields-def}--\ref{tab:chap3-correlation-matrix} report variable descriptions, summary statistics, and the correlation matrix. Summary statistics are presented on the original scale after 1\% winsorization, including raw WEPA projection scores. The correlation matrix is computed using winsorized and standardized data.

\begin{table}[!htb]
\footnotesize
\caption{Variable descriptions.}
\label{tab:chap3-dataset-fields-def}
\begin{adjustwidth}{-\extralength}{0cm}
\renewcommand{\arraystretch}{1.0}
\begin{tabularx}{\fulllength}{>{\raggedright\arraybackslash}p{2.8cm}X}
\toprule
\textbf{Variables} & \textbf{Description} \\
\midrule

\multicolumn{2}{l}{\textbf{Panel Identifiers 
}} \\
userid       & De-identified platform ID, used for individual fixed effects. \\
WeekI        & Number of weeks between the current observation week and the user's registration week. \\

\midrule
\multicolumn{2}{l}{\textbf{Dependent Variable}} \\
ExDur & Log-transformed weekly exercise duration in minutes, computed as $\ln(x+1)$. \\

\midrule
\multicolumn{2}{l}{\textbf{Independent Variables (WEPA projection scores)}} \\
GoalCommit   & Projection score of the weekly text vector onto the Goal Commitment semantic axis. \\
GoalSpec     & Projection score of the weekly text vector onto the Goal Specificity semantic axis. \\
GoalDiff     & Projection score of the weekly text vector onto the Goal Difficulty semantic axis. \\
SelfEfficacy & Sum of the four subdimensional projection scores: MastExp + VicExp + SocPer + PhyState. \\
\quad MastExp    & Projection score of the weekly text vector onto the Mastery Experience semantic axis. \\
\quad VicExp     & Projection score of the weekly text vector onto the Vicarious Experience semantic axis. \\
\quad SocPer     & Projection score of the weekly text vector onto the Social Persuasion semantic axis. \\
\quad PhyState   & Projection score of the weekly text vector onto the Physiological States semantic axis. \\

\midrule
\multicolumn{2}{l}{\textbf{Time-Varying Control Variables}} \\
age          & Age in years, calculated as the difference between the observation week and the user's date of birth. \\
StrLen & Character length of the user--week text, computed as the number of characters in the text string. \\
{ValidStrLen} & {Effective length of the cleaned user--week text, measured by valid tokens for WEPA scoring.} \\
DurChg       & First difference in logged exercise duration between adjacent weeks: $ExDur_{i,t} - ExDur_{i,t-1}$. \\
SocEngage   & Log-transformed weekly social engagement, computed as $\ln(x+1)$ from discussion and check-in activities. \\
SocFeedback & Log-transformed weekly social feedback, computed as $\ln(x+1)$ from likes and comments received. \\
\bottomrule
\end{tabularx}\end{adjustwidth}
\noindent{\footnotesize{\textit{{Note: 
}}WEPA variables are raw scalar projection scores. Continuous predictors are standardized in regression analysis. {ValidStrLen is used only in the textual-sparsity sensitivity analysis.}}}

\end{table}

\vspace{-9pt}
\begin{table}[!htb]
\small
\caption{Summary statistics.}
\label{tab:chap3-dataset-stats}
\begin{adjustwidth}{-\extralength}{0cm}
\renewcommand{\arraystretch}{1.0}
\begin{tabularx}{\fulllength}{>{\raggedright\arraybackslash}p{2.8cm}>{\centering\arraybackslash}X>{\centering\arraybackslash}X>{\centering\arraybackslash}X>{\centering\arraybackslash}X>{\centering\arraybackslash}X>{\centering\arraybackslash}X}
\toprule
\textbf{Variable} & \textbf{Users} & \textbf{Obs} & \textbf{Mean} & \textbf{S.D.} & \textbf{Min} & \textbf{Max} \\
\midrule
WeekI        & 177,829 & 2,668,298 & 97.636 & 74.244 & 2.000 & 338.000 \\
ExDur        & 177,829 & 2,668,298 & 2.514 & 2.343 & 0.000 & 8.504 \\
SelfEfficacy & 177,829 & 2,668,298 & 6.680 & 2.006 & 3.616 & 12.337 \\
\quad MastExp & 177,829 & 2,668,298 & 1.950 & 0.608 & 1.072 & 3.755 \\
\quad VicExp  & 177,829 & 2,668,298 & 1.175 & 0.369 & 0.608 & 3.999 \\
\quad SocPer  & 177,829 & 2,668,298 & 1.929 & 0.625 & 1.049 & 3.989 \\
\quad PhyState & 177,829 & 2,668,298 & 1.626 & 0.500 & 0.887 & 3.535 \\
GoalCommit   & 177,829 & 2,668,298 & 1.101 & 0.484 & 0.403 & 3.291 \\
GoalSpec     & 177,829 & 2,668,298 & $-$0.354 & 0.408 & $-$0.926 & 2.656 \\
GoalDiff     & 177,829 & 2,668,298 & $-$1.079 & 0.404 & $-$1.623 & 2.124 \\
age          & 177,829 & 2,668,298 & 29.459 & 8.002 & 10.003 & 80.000 \\
StrLen       & 177,829 & 2,668,298 & 50.559 & 58.858 & 1.000 & 188.000 \\
{ValidStrLen}  & {177,829} & {2,668,298} & {44.198} & {51.364} & {0.000} & {164.000}\\
DurChg       & 177,829 & 2,668,298 & 0.482 & 1.458 & $-$8.368 & 8.349 \\
SocEngage    & 177,829 & 2,668,298 & 1.811 & 0.820 & 0.693 & 3.258 \\
SocFeedback  & 177,829 & 2,668,298 & 2.646 & 1.757 & 0.000 & 5.501 \\
\bottomrule
\end{tabularx}
\end{adjustwidth}
\end{table}

\clearpage
\begin{landscape}
\begin{table}[p]
\centering
\scriptsize
\caption{Correlation matrix.}
\label{tab:chap3-correlation-matrix}
\renewcommand{\arraystretch}{1.45}
\setlength{\tabcolsep}{5pt}
\resizebox{0.96\linewidth}{!}{%
\begin{tabular}{@{}l*{14}{r}@{}}
\toprule
\textbf{Variable} & \textbf{(1)} & \textbf{(2)} & \textbf{(3)} & \textbf{(4)} & \textbf{(5)} & \textbf{(6)} & \textbf{(7)} & \textbf{(8)} & \textbf{(9)} & \textbf{(10)} & \textbf{(11)} & \textbf{(12)} & \textbf{(13)} & \textbf{(14)} \\
\midrule
(1) ExDur            & 1.000 &       &       &       &       &       &       &       &       &       &       &       &       &       \\
(2) GoalCommit        & $-$0.053 & 1.000 &       &       &       &       &       &       &       &       &       &       &       &       \\
(3) GoalSpec         & $-$0.072 & 0.084 & 1.000 &       &       &       &       &       &       &       &       &       &       &       \\
(4) GoalDiff         & $-$0.025 & $-$0.544 & 0.147 & 1.000 &       &       &       &       &       &       &       &       &       &       \\
(5) SelfEfficacy    & $-$0.030 & 0.862 & $-$0.171 & $-$0.669 & 1.000 &       &       &       &       &       &       &       &       &       \\
(6) MastExp          & $-$0.005 & 0.831 & $-$0.197 & $-$0.647 & 0.977 & 1.000 &       &       &       &       &       &       &       &       \\
(7) VicExp           & $-$0.002 & 0.798 & $-$0.085 & $-$0.677 & 0.918 & 0.858 & 1.000 &       &       &       &       &       &       &       \\
(8) SocPer           & $-$0.035 & 0.803 & $-$0.222 & $-$0.601 & 0.965 & 0.957 & 0.834 & 1.000 &       &       &       &       &       &       \\
(9) PhyState        & $-$0.071 & 0.841 & $-$0.146 & $-$0.615 & 0.937 & 0.894 & 0.792 & 0.872 & 1.000 &       &       &       &       &       \\
(10) age            & 0.045 & 0.036 & 0.050 & $-$0.007 & 0.026 & 0.024 & 0.015 & 0.024 & 0.036 & 1.000 &       &       &       &       \\
(11) StrLen         & 0.308 & $-$0.074 & 0.092 & 0.045 & $-$0.172 & $-$0.179 & $-$0.149 & $-$0.163 & $-$0.162 & 0.055 & 1.000 &       &       &       \\
(12) DurChg          & 0.333 & $-$0.021 & $-$0.066 & $-$0.025 & 0.011 & 0.018 & 0.011 & 0.013 & $-$0.002 & 0.003 & 0.069 & 1.000 &       &       \\
(13) SocEngage       & 0.696 & $-$0.011 & 0.044 & $-$0.021 & $-$0.045 & $-$0.043 & $-$0.033 & $-$0.047 & $-$0.048 & 0.129 & 0.507 & 0.218 & 1.000 &       \\
(14) SocFeedback    & 0.510 & $-$0.128 & 0.047 & 0.114 &$-$0.185 & $-$0.193 & $-$0.185 & $-$0.176 & $-$0.147 & 0.120 & 0.653 & 0.145 & 0.756 & 1.000 \\
\bottomrule
\end{tabular}
}
\end{table}
\end{landscape}
\clearpage

The correlation matrix shows several notable relationships. Goal commitment is negatively correlated with goal difficulty ($r=-0.544$), indicating that more difficulty-related expression tends to coincide with lower commitment in this sample. {The four self-efficacy dimensions are highly correlated ($r=0.79$--$0.96$), suggesting substantial shared variance}. Social engagement is strongly correlated with exercise duration ($r=0.696$), supporting the inclusion of social controls in the regression models.

\section{Validation Studies}
\subsection{Study I: Measurement Foundation and Benchmark Validation}
\label{subsec:study1_measurement_foundation}

Study I establishes the measurement foundation of WEPA through anchor-word development, anchor-word validation, benchmark-based evaluation against human annotations, and cross-temporal stability assessment. Because WEPA generates a single construct score for each user--week observation and the focal constructs are expected to vary over time, classical reliability measures such as internal consistency and test--retest reliability are not directly applicable. We therefore examine whether the anchor-word dictionaries exhibit the expected semantic structure, whether WEPA scores align with human judgments and outperform a dictionary-based baseline, and whether semantic axes retain stable orientation across the observation period.

\subsubsection{Anchor-Word Development and Validation}
\label{subsubsec:anchor_word_development}

We identified focal psychological constructs from Goal-Setting Theory and Social Cognitive Theory. The goal-setting constructs include goal commitment, goal specificity, and goal difficulty. We also examine self-efficacy and its four source dimensions, including mastery experience, vicarious experience, social persuasion, and physiological states. These constructs cover both unidimensional and multidimensional structures, are theoretically central to exercise behavior, and can be meaningfully represented through bipolar \mbox{semantic dimensions}.

In the Chinese fitness domain, anchor words were selected to capture theoretically meaningful semantic contrasts. Goal specificity contrasts quantified goal indicators with vague outcome expectations. Goal difficulty contrasts high-effort or challenge-related expressions with low-intensity or easy activities. Goal commitment contrasts determination and follow-through with hesitation, delay, or disengagement. For self-efficacy, mastery experience contrasts success with failure-related expressions, vicarious experience contrasts learning from others with unguided trial-and-error, social persuasion contrasts supportive with discouraging feedback, and physiological states contrast energetic or positive conditions with fatigued or depleted ones.

{Some dimensions require context-specific interpretation. The goal difficulty axis captures difficulty-related expressions in online fitness discourse, including perceived obstacles, fatigue, and challenge-related language, and should not be treated as a direct measure of objective goal difficulty in the original Goal-Setting Theory. Similarly, the social persuasion axis captures platform expressions related to encouragement or discouragement, whose behavioral meaning may depend on whether they reflect stable support or compensatory help-seeking.}

Table~\ref{tab:anchor_word_glosses} presents representative English glosses of selected Chinese anchor words to help international readers to understand their semantic content and polarity. The empirical analysis uses the original Chinese anchor words.

\begin{table}[!htb]

\caption{English glosses of selected chinese anchor words.}
\label{tab:anchor_word_glosses}
\begin{adjustwidth}{-\extralength}{0cm}
\footnotesize
\renewcommand{\arraystretch}{1.1}
\begin{tabularx}{\fulllength}{@{}>{\raggedright\arraybackslash}p{3.2cm}
>{\raggedright\arraybackslash}p{1.5cm}
X@{}}
\toprule
\textbf{Construct} & \textbf{Direction} & \textbf{Anchor-Word Glosses} \\
\midrule
Goal Specificity & Positive & kg; BMI; pace; minutes; calories; heart rate \\
Goal Specificity & Negative & lose weight; get stronger; look better; look attractive; feel better \\
Goal Difficulty & Positive & difficult; grit one's teeth; tearing; breakthrough; overcome; limit \\
Goal Difficulty & Negative & basic; beginner; entry-level; activation; soothing; easy \\
Goal Commitment & Positive & determination; firm resolve; persistence; plan; follow through; deliver \\
Goal Commitment & Negative & delay; hesitation; interruption; termination; retreat \\
Mastery Experience & Positive & success; achievement; victory; reach the target; complete \\
Mastery Experience & Negative & failure; give up; lose control; stall; come to an end \\
Vicarious Experience & Positive & seek help; ask for advice; consult; expert; senior; imitate; refer to; template \\
Vicarious Experience & Negative & alone; no one to teach; rely on feeling; grope one's way; fool around; act blindly \\
Social Persuasion & Positive & recognition; support; affirmation; encouragement; praise; compliment; trust \\
Social Persuasion & Negative & denial; doubt; ridicule; sarcasm; mockery; neglect; feel disheartened \\
Physiological State & Positive & vitality; energy; happiness; confidence; fully energized; alert; high-spirited \\
Physiological State & Negative & weakness; drowsiness; fatigue; stress; self-abasement; listlessness \\
\bottomrule
\end{tabularx}\end{adjustwidth}

\noindent{\footnotesize \textit{Note:} The English entries are interpretive glosses provided for readability and are not used as operational dictionaries in the empirical analysis. Where necessary, contextualized phrases rather than literal translations are used to preserve construct meaning.}

\end{table}

The anchor-word dictionaries were developed in three stages. First, theory pooling drew on Goal-Setting Theory, Social Cognitive Theory, and fitness-community language to generate 260 candidate words. After screening against the 200-dimensional GloVe vocabulary trained on the 816 MB corpus, 
251 valid candidates remained. Second, three eHealth scholars independently evaluated the fit between candidate words and construct definitions. Words were retained when at least two experts agreed on their assigned category. Inter-rater agreement was high, with Krippendorff's $\alpha = 0.8249$, and 232 theoretically relevant anchor words were retained.

{Finally, we conducted semantic-space diagnostics. For each construct, we calculated average cosine similarity within the positive pole, within the negative pole, and between poles. This diagnostic assessed whether words within the same pole were, on average, closer to one another than to words from the opposite pole. Table~\ref{tab:anchor_semantic_validation} shows that all constructs follow this expected pattern. Although the absolute cosine values are moderate, this is plausible because anchor words are not intended to be strict synonyms. They function as theoretically related expressions aligned with the same pole, supporting subsequent semantic-axis construction.}

\begin{table}[!htb]
\caption{Semantic-space diagnostic results for anchor-word sets.}
\label{tab:anchor_semantic_validation}
\begin{adjustwidth}{-\extralength}{0cm}
\setlength{\cellWidtha}{\fulllength/4-2\tabcolsep-0in}
\setlength{\cellWidthb}{\fulllength/4-2\tabcolsep-0in}
\setlength{\cellWidthc}{\fulllength/4-2\tabcolsep-0in}
\setlength{\cellWidthd}{\fulllength/4-2\tabcolsep-0in}

\scalebox{1}[1]{\begin{tabularx}{\fulllength}{>{\raggedright\arraybackslash}m{\cellWidtha}>{\centering\arraybackslash}m{\cellWidthb}>{\centering\arraybackslash}m{\cellWidthc}>{\centering\arraybackslash}m{\cellWidthd}}
\toprule
    \textbf{Construct} & \textbf{Average Within-POS Similarity} & \textbf{Average Within-NEG Similarity} & \textbf{Average Between-Pole Similarity} \\
    \midrule
    Goal Specificity      & 0.3399 & 0.3216 & 0.2282 \\
    Goal Commitment       & 0.2095 & 0.1586 & 0.0899 \\
    Vicarious Experience  & 0.1967 & 0.0723 & 0.0426 \\
    Mastery Experience    & 0.4068 & 0.2077 & 0.1763 \\
    Physiological State   & 0.3086 & 0.2549 & 0.1524 \\
    Social Persuasion     & 0.2930 & 0.1505 & 0.0834 \\
    Goal Difficulty       & 0.3054 & 0.4454 & 0.2968 \\
    \bottomrule
\end{tabularx}}\end{adjustwidth}

\noindent{\footnotesize{\textit{Note:} The table reports average cosine similarities for each construct in the embedding space. When both within-pole similarities exceed the between-pole similarity, the anchor-word set is considered to show the expected pattern of separation between poles.}}

\end{table}

\subsubsection{Human Benchmark and Algorithm Comparison}
\label{subsubsec:benchmark_comparison}

{To evaluate WEPA against human benchmarks, we selected two theoretically distinct constructs. Goal specificity represents a relatively concrete cognitive--behavioral dimension of Goal-Setting Theory and tests whether WEPA can recover quantified goal statements. Physiological states represent an affective--somatic source of self-efficacy and provide a more difficult test because expressions such as \textit{energized} or \textit{exhausted} are often implicit and fragmented in short social-media text.}

{We constructed two annotated datasets from the Keep corpus. The goal-specificity benchmark includes 518 posts with quantifiable goals, and the physiological-states benchmark includes 536 posts with identifiable bodily or emotional descriptions. All texts contain at least 10 characters. The same three eHealth scholars annotated both datasets. They identified quantifiable behavioral indicators for goal specificity and physiological or affective indicators for physiological states without inferring broader motivational meanings. Inter-annotator agreement was high for both constructs, with Cronbach's $\alpha = 0.965$ for goal specificity and 0.922 for physiological states.}

{We use a dictionary-based method as a transparent closed-vocabulary baseline. Both methods rely on the same anchor-word sets, but their scoring logic differs. The dictionary method relies on exact keyword matching, whereas WEPA uses distributed semantic aggregation. This baseline is not intended to represent all possible dictionary systems. It provides a conservative reference for evaluating whether semantic projection improves coverage and agreement with human judgments under short-text conditions.}

The dictionary-based method computes scores using the normalized difference in anchor-word frequencies \citep{Eichstaedt2020ClosedOpen}:
\begin{equation}
dict\_score = \frac{ |\mathcal{W} \cap A_{\text{pos}}| - |\mathcal{W} \cap A_{\text{neg}}| }{ |\mathcal{W} \cap A_{\text{pos}}| + |\mathcal{W} \cap A_{\text{neg}}| + 1 }
\label{eq:dict}
\end{equation}
where $\mathcal{W}$ denotes the set of words in the focal text, and $A_{\text{pos}}$ and $A_{\text{neg}}$ denote the positive and negative anchor-word sets. Stronger agreement with human annotations indicates a better capture of the intended construct as expressed in the text.

Table~\ref{tab:reliability_validation} reports the benchmark comparison. For goal specificity, WEPA achieves a Spearman rank correlation of $\rho = 0.895$ ($p < 0.001$), higher than the dictionary baseline ($\rho = 0.733$, $p < 0.001$). {For physiological states, the contrast is larger. The dictionary method covers only 52 of 536 texts (9.7\%) and shows weak agreement with human judgments ($\rho = 0.161$, $p < 0.001$), whereas WEPA scores all texts and achieves strong agreement ($\rho = 0.850$, $p < 0.001$).}

\begin{table}[!htb]\small
\caption{Benchmark Comparison: WEPA vs. dictionary-based method.}
\label{tab:reliability_validation}
\setlength{\cellWidtha}{\textwidth/5-2\tabcolsep+0.6in}
\setlength{\cellWidthb}{\textwidth/5-2\tabcolsep-0.3in}
\setlength{\cellWidthc}{\textwidth/5-2\tabcolsep-0in}
\setlength{\cellWidthd}{\textwidth/5-2\tabcolsep-0.3in}
\setlength{\cellWidthe}{\textwidth/5-2\tabcolsep-0in}
\scalebox{1}[1]{\begin{tabularx}{\textwidth}{>{\raggedright\arraybackslash}m{\cellWidtha}>{\centering\arraybackslash}m{\cellWidthb}>{\centering\arraybackslash}m{\cellWidthc}>{\centering\arraybackslash}m{\cellWidthd}>{\centering\arraybackslash}m{\cellWidthe}}
\toprule
    \multirow{2.5}{*}{\textbf{Measurement Method}} & \multicolumn{2}{c}{\textbf{Goal Specificity}} & \multicolumn{2}{c}{\textbf{Physiological State}} \\
    \cmidrule{2-5}
    & $\boldsymbol{\rho}$ & \textbf{Coverage} & $\boldsymbol{\rho}$ & \textbf{Coverage} \\
    \midrule
    Dictionary-based & 0.733 & 32.4\% & 0.161 & 9.7\% \\
    WEPA  & \textbf{0.895 
} & \textbf{100\%} & \textbf{0.850} & \textbf{100\%} \\
    \bottomrule
\end{tabularx}}

\noindent{\footnotesize{\textit{Note:} N = 518 (goal specificity) and 536 (physiological state) human-annotated texts. Coverage denotes the percentage of texts that receive a non-missing score.}}

\end{table}

These results suggest that semantic projection improves both coverage and benchmark agreement under realistic short-text conditions. The advantage is especially clear for constructs expressed through implicit, affective, or fragmented language, where exact lexical matching often fails. {At the same time, the benchmark should not be interpreted as direct human-annotation validation for all seven constructs. It provides evidence across two theoretically distinct cases, one relatively concrete cognitive--behavioral construct and one affective--somatic construct.}

\subsubsection{{Cross-Temporal Stability of Semantic Axes}}
\label{subsubsec:cross-temporal-stability-of-axis}

{The benchmark comparison establishes that WEPA captures the intended constructs in human-annotated texts. For longitudinal measurement, it is also necessary to examine whether semantic axes retain stable orientation over time. If axis orientation shifts because of linguistic evolution or corpus-specific estimation variance, observed score changes may partly reflect measurement artifacts.}

{To assess cross-temporal robustness, we distinguish between axis orientation and semantic structure. We partition the full corpus by calendar year from 2015 to 2021 and train separate yearly embedding models using the same hyperparameters as the main specification. Each yearly embedding space is aligned to the full-period reference space using a Procrustes transformation, so that all yearly models share a common coordinate system \citep{Yao2017Dynamic}. }

{The full-period embedding is used as the reference because it provides a more stable estimate of the semantic space than yearly embeddings, especially in early sparse periods, and avoids anchoring the analysis to any single-year realization \citep{Cassani2021Words}. This choice does not introduce forward-looking bias because the analysis does not involve prediction. The full-period embedding serves only as a technical reference frame for comparing semantic representations across time.}

{For each construct, we compute two indicators. Cosine similarity between the yearly semantic axis and the full-period axis captures axis stability, or whether the overall direction of the construct axis remains stable. Spearman rank correlation captures rank-order consistency, or whether the relative ordering of anchor words along the construct dimension is preserved. We place less emphasis on 2015 because the Keep platform was launched in February 2015 and the first-year corpus is sparse, approximately 23 MB, compared with \mbox{134 MB} in 2016.}

\begin{figure}[!htbp]
\begin{adjustwidth}{-\extralength}{0cm}
\centering
\includegraphics[width=\fulllength]{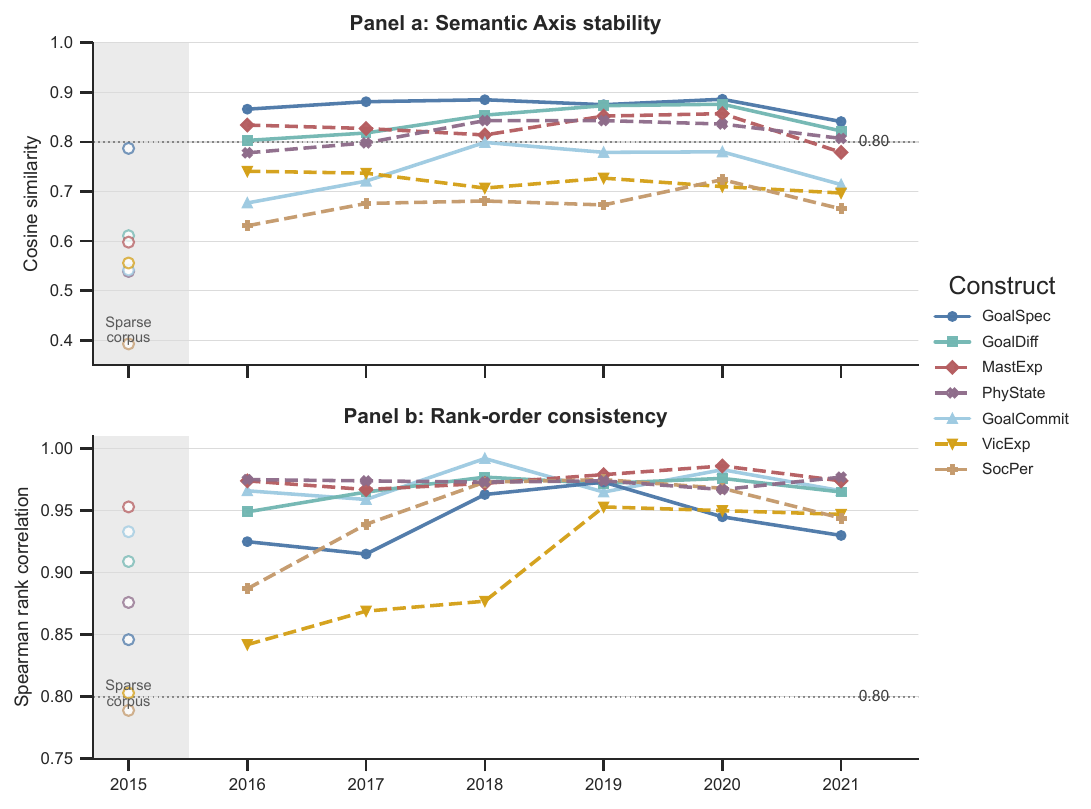}
\end{adjustwidth}
\caption{{Cross-temporal stability of semantic axes. 
Panel (\textbf{a}) shows semantic axis stability based on cosine similarity. 
Panel (\textbf{b}) shows rank-order consistency based on Spearman correlations.}}
\label{fig:temporal_axis_stability}
\end{figure}

{Figure~\ref{fig:temporal_axis_stability} presents the results. Panel (a) reports semantic axis stability based on cosine similarity, and Panel (b) reports rank-order consistency based on Spearman correlations.}

{These two indicators capture complementary aspects of temporal stability. Axis stability reflects whether the overall geometric direction is preserved, while rank-order consistency reflects whether the internal semantic structure of anchor words remains stable. The results show that axis orientation varies across constructs, but rank-order consistency remains consistently high after 2016. This pattern suggests that WEPA provides a stable semantic measurement structure over time, especially at the level of internal semantic ordering. The analysis supports the longitudinal interpretability of WEPA scores, although it should not be treated as a full proof of measurement invariance across platforms \mbox{or contexts}.}\vspace{-3pt}

\subsection{Study II: Criterion Validity of Goal-Setting Constructs}
\label{subsec:study2_goal_setting}

Study II evaluates whether WEPA-derived scores for goal commitment, goal specificity, and goal difficulty are associated with subsequent exercise behavior in theoretically consistent directions.

\subsubsection{Hypotheses
}

Goal-Setting Theory argues that clear, specific, and appropriately challenging goals, when combined with feedback, can improve individual performance \citep{Locke2019The}. Building on this logic, we develop expectations for three goal-setting constructs. Because WEPA measures these constructs through users' own expressions in platform text, the hypotheses concern text-based indicators of goal-related psychological states rather than experimentally assigned goal conditions.

Goal commitment reflects persistence, motivational investment, and determination. Users with higher commitment are more likely to sustain exercise behavior \citep{Brandstatter2021Persistence}. Goal specificity refers to quantified and actionable goal statements, such as \textit{run 5 km per week}. Specific goals reduce behavioral ambiguity and provide clearer implementation paths \citep{Wirth2009The}. Accordingly, WEPA goal commitment and goal specificity scores are expected to be positively associated with future exercise duration.

{Goal difficulty requires a context-specific interpretation. In classical Goal-Setting Theory, difficult goals can promote performance when they are accepted and supported by sufficient ability and resources \citep{Senko2005Achievement}. In online fitness discourse, however, expressions such as \textit{too hard}, \textit{I cannot keep going}, or \textit{I am exhausted} may signal perceived obstacles, fatigue, or frustration. The WEPA goal difficulty score is therefore interpreted as expressed difficulty in platform text, and we expect it to be negatively associated with subsequent exercise duration.}

Based on these arguments, we propose the following hypotheses:

\begin{Hypothesis}
The WEPA goal commitment score is positively associated with future exercise duration.
\end{Hypothesis}

\begin{Hypothesis}
The WEPA goal specificity score is positively associated with future exercise duration.
\end{Hypothesis}

\begin{Hypothesis}
\textls[-15]{The WEPA goal difficulty score is negatively associated with future exercise duration.}
\end{Hypothesis}

\subsubsection{Model Specification}

The three goal-setting constructs are derived from users' weekly text using WEPA and measured as continuous construct scores. Because goal commitment and goal difficulty are conceptually related and empirically correlated ($r = -0.544$; see Table~\ref{tab:chap3-correlation-matrix}), estimating all three constructs in a joint model would complicate interpretation. Joint estimation would yield partial effects conditional on other constructs, which is less aligned with the goal of assessing criterion validity for each construct individually. We therefore estimate separate individual fixed-effects panel models for each construct.

Multicollinearity diagnostics indicate that joint estimation is statistically feasible, with maximum condition indices below 2.07 and variance inflation factors below 1.57. The use of separate models is therefore motivated by interpretability rather than by multicollinearity~concerns.

Using individual fixed-effects models, we estimate the following regression for each~construct:
\begin{equation}
ExDur_{i,t+1} = \alpha + \beta_k \cdot Construct_{k,i,t} + \gamma \cdot Controls_{i,t} + \mu_i + \epsilon_{i,t}, \quad k \in \{1,2,3\}
\label{eq:study2_separate_models}
\end{equation}
where $ExDur_{i,t+1}$ is log-transformed exercise duration at time $t+1$, and $Construct_{k,i,t}$ denotes goal commitment, goal specificity, or goal difficulty at time $t$. $\mu_i$ represents individual fixed effects, and standard errors are clustered at the user level. The time-varying controls include age, weekly text length, exercise-duration change, social engagement, and \mbox{social feedback}.

{Criterion validity is assessed through temporally ordered predictive associations between construct scores at time $t$ and exercise behavior at time $t+1$. Prediction refers to cross-temporal statistical association, not causal prediction or intervention effects. Weekly aggregation limits temporal precision because users may exercise or post at any point within a week. Directional consistency between construct scores and the external behavioral criterion provides evidence for criterion validity within this observational framework.}

\subsubsection{Results
}

Table~\ref{tab:chap3-unidim-validation-results} reports the criterion-validity results. Goal commitment is positively associated with exercise duration at time $t+1$ ($\beta = 0.069$, $p < 0.001$), supporting H1. Goal specificity is also positively associated with exercise duration ($\beta = 0.019$, $p < 0.001$), supporting H2. Goal difficulty is negatively associated with exercise duration ($\beta = -0.053$, $p < 0.001$), supporting H3.

{Because the dependent variable is log-transformed, the coefficients can be interpreted approximately as semi-elasticities. A one-standard-deviation increase in goal commitment, goal specificity, and goal difficulty corresponds to about +7.1\%, +1.9\%, and $-$5.2\%, respectively, in exercise duration at $t+1$. Using the raw mean of weekly exercise duration before log transformation as a reference (15.61 minutes/week), these effects translate into about +1.12, +0.30, and $-$0.81 minutes/week. These modest raw-minute equivalents should be interpreted as statistically reliable behavioral associations, not as large individual-level intervention effects.}

These findings provide criterion-validity evidence for WEPA in measuring unidimensional psychological constructs.

\begin{table}[!htb]
\caption{Criterion validity of goal-setting constructs.}
\label{tab:chap3-unidim-validation-results}
\begin{adjustwidth}{-\extralength}{0cm}
\setlength{\cellWidtha}{\fulllength/5-2\tabcolsep-0in}
\setlength{\cellWidthb}{\fulllength/5-2\tabcolsep-0in}
\setlength{\cellWidthc}{\fulllength/5-2\tabcolsep-0in}
\setlength{\cellWidthd}{\fulllength/5-2\tabcolsep-0in}
\setlength{\cellWidthe}{\fulllength/5-2\tabcolsep-0in}
\scalebox{1}[1]{\begin{tabularx}{\fulllength}{>{\raggedright\arraybackslash}m{\cellWidtha}>{\centering\arraybackslash}m{\cellWidthb}>{\centering\arraybackslash}m{\cellWidthc}>{\centering\arraybackslash}m{\cellWidthd}>{\centering\arraybackslash}m{\cellWidthe}}
\toprule
  & \multicolumn{4}{c}{\textbf{DV:} $\boldsymbol{ExDur_{i,t+1}}$} \\
  \cmidrule{2-5}
  & \textbf{(1) Baseline Model} & \textbf{(2) GoalCommit} & \textbf{(3) GoalSpec} & \textbf{(4) GoalDiff} \\
  \midrule
      GoalCommit  & & \textbf{0.069 
} *** & &  \\
                  & & \textbf{(0.002)} & &  \\
      GoalSpec    & & & \textbf{0.019} *** &  \\
                  & & & \textbf{(0.002)} &  \\
      GoalDiff    & & & & \textbf{$-$0.053} *** \\
                  & & & & \textbf{(0.002)} \\
      age         & $-$0.942 *** & $-$0.945 *** & $-$0.944 *** & $-$0.942 *** \\
                  & (0.004) & (0.004) & (0.004) & (0.004)  \\
      StrLen      & 0.011 *** & 0.011 *** & 0.010 *** & 0.011 ***  \\
                  & (0.002) & (0.002) & (0.002) & (0.002)  \\
      DurChg      & 0.066 *** & 0.066 *** & 0.066 *** & 0.065 ***  \\
                  & (0.001) & (0.001) & (0.001) & (0.001)  \\
      SocEngage & 0.685 *** & 0.681 *** & 0.685 *** & 0.682 ***  \\
                  & (0.004) & (0.004) & (0.004) & (0.004)  \\
      SocFeedback  & $-$0.059 *** & $-$0.054 *** & $-$0.058 *** & $-$0.055 *** \\
                  & (0.002) & (0.002) & (0.002) & (0.002)  \\
     \midrule
     Ind FE            & Yes & Yes & Yes & Yes \\
     Observations      & 2,567,730 & 2,567,730 & 2,567,730 & 2,567,730  \\
     N. of Users       & 175,114 & 175,114 & 175,114 & 175,114 \\
     Within $R^2$      & 0.336  & 0.336 & 0.336 & 0.336  \\
  \bottomrule
\end{tabularx}}\end{adjustwidth}
\noindent{\footnotesize{\textit{Note:} Standard errors in parentheses and clustered at the user level. 
*** \emph{p} < 0.001.}}

\end{table}

\subsection{Study III: Criterion Validity of Self-Efficacy Dimensions}
\label{subsec:study3_self_efficacy}
Study III extends the criterion-validity assessment to multidimensional measurement. It examines whether WEPA-derived scores for overall self-efficacy and its four source dimensions are associated with subsequent exercise behavior in theoretically \mbox{expected directions}.

\subsubsection{Hypotheses}
According to Social Cognitive Theory, self-efficacy is shaped by four sources of information, including mastery experience, vicarious experience, social persuasion, and physiological states \citep{Bandura1977SelfEfficacy}. This four-dimensional structure makes self-efficacy a useful case for testing whether WEPA can distinguish behavioral associations among dimensions within the same construct. Because these dimensions are measured through linguistic expressions in platform text, the hypotheses concern WEPA-derived textual indicators rather than a direct experimental manipulation of self-efficacy sources.

Mastery experience reflects successful task completion and is widely regarded as the strongest source of self-efficacy. Vicarious experience arises from observing similar others succeed and provides behavioral models for action. Expressions of prior accomplishment or learning from others should therefore be positively associated with future exercise duration. Physiological states reflect subjective evaluations of physical condition and emotional arousal. Positive bodily and emotional states may support behavioral maintenance, while fatigue or negative affect may weaken efficacy beliefs. Thus, the WEPA physiological states score is also expected to be positively associated with future exercise duration.

{Social persuasion requires a context-specific interpretation in digital fitness communities. In Bandura's original theory, persuasive encouragement from others can strengthen efficacy beliefs. In platform discourse, however, references to external affirmation, such as, "\textit{I need likes 
}", "\textit{please encourage me}", or "\textit{I need support to continue}", may appear as compensatory expressions during motivational blockage. We therefore expect the WEPA social persuasion score to be negatively associated with subsequent exercise duration in this platform context. This hypothesis reflects the user-expressed salience of persuasion- or support-related language and should not be interpreted as a general claim that social persuasion weakens self-efficacy.}

Based on these arguments, we propose the following hypotheses.

\begin{Hypothesis}
The WEPA self-efficacy score is positively associated with future exercise duration.
\end{Hypothesis}

\begin{Hypothesis}
\textls[15]{The WEPA mastery experience score is positively associated with future exercise~duration.}
\end{Hypothesis}

\begin{Hypothesis}
\textls[15]{The WEPA vicarious experience score is positively associated with future exercise~duration.}
\end{Hypothesis}

\begin{Hypothesis}
\textls[15]{The WEPA physiological states score is positively associated with future exercise~duration.}
\end{Hypothesis}

\begin{Hypothesis}
\textls[15]{The WEPA social persuasion score is negatively associated with future exercise~duration.}
\end{Hypothesis}

\subsubsection{Model Specification}
Study III adopts the joint model as the primary specification because the four source dimensions represent theoretically distinct components of the same multidimensional construct. The goal is to assess each dimension's association with subsequent behavior and to examine its relative contribution after accounting for the other sources. This logic is consistent with the theoretical role of self-efficacy sources in Social Cognitive Theory.

Confirmatory factor analysis results support this multidimensional structure\linebreak (CFI = 0.999, TLI = 0.998, RMSEA = 0.038). {Although the four dimensions are highly correlated ($r = 0.79$--$0.96$), the joint model remains informative because it captures conditional associations among theoretically related components. Multicollinearity diagnostics indicate that the model is estimable. The VIFs for MastExp and SocPer are 13.09 and 10.31, slightly above the conventional threshold of 10, while the maximum condition index is 8.56, well below the critical threshold of 30 \citep{O’Brien2007A, Kalnins2023The}. These diagnostics call for a cautious interpretation of coefficient magnitudes without invalidating the joint specification.}

Using an individual fixed-effects model, the joint specification is written as follows:
\begin{equation}
ExDur_{i,t+1} = \alpha + \sum_{m=1}^{4} \beta_m \cdot Dimension_{m,i,t} + \gamma \cdot Controls_{i,t} + \mu_i + \epsilon_{i,t},
\label{eq:study3_joint_model}
\end{equation}
where $Dimension_{m,i,t}$ denotes mastery experience, vicarious experience, social persuasion, or physiological states at time $t$. The joint model estimates each source dimension conditional on the other three dimensions.

{As a sensitivity check, we also estimate separate single-dimension models:}
\begin{equation}
{ExDur_{i,t+1} = \alpha + \beta_m \cdot Dimension_{m,i,t} + \gamma \cdot Controls_{i,t} + \mu_i + \epsilon_{i,t}},
\quad {m \in \{1,2,3,4\}.}
\label{eq:study3_separate_model}
\end{equation}

{The separate models estimate gross associations between each source expression and subsequent exercise behavior, while the joint model estimates conditional associations.} This comparison helps to assess whether the joint estimates are sensitive to intercorrelations among the four dimensions. The controls, fixed effects, and clustered standard errors follow Study II.

\subsubsection{Results}

Table~\ref{tab:chap3-multidim-validation-results} reports the criterion-validity results for overall self-efficacy and its four source dimensions. All models use the same analysis sample. The overall self-efficacy score is positively associated with exercise duration at time $t+1$ ($\beta = 0.008$, $p < 0.001$), \mbox{supporting H4}.

\begin{table}[!htb]
\caption{Criterion validity of self-efficacy dimensions.}
\label{tab:chap3-multidim-validation-results}
\begin{adjustwidth}{-\extralength}{0cm}
\centering
\renewcommand{\arraystretch}{1.0}
\footnotesize
\begin{tabularx}{\fulllength}{@{}
  >{\raggedright\arraybackslash}p{2.7cm}
  >{\centering\arraybackslash}X
  >{\centering\arraybackslash}X
  >{\centering\arraybackslash}X
  >{\centering\arraybackslash}X
  >{\centering\arraybackslash}X
  >{\centering\arraybackslash}X
  >{\centering\arraybackslash}X @{}}
  \toprule
  & \multicolumn{7}{c}{\textbf{DV:} $\boldsymbol{ExDur_{i,t+1}}$} \\
  \cmidrule{2-8}

  & 
  & 
  & \multicolumn{4}{c}{\textbf{Separate Models}} 
  & \textbf{Joint Model} \\
  \cmidrule{4-8}

  & \textbf{(1) Baseline }
  & \textbf{(2) Overall SE} 
  & \textbf{(3) MastExp} 
  & \textbf{(4) VicExp }
  & \textbf{(5) SocPer }
  & \textbf{(6) PhyState }
  & \textbf{(7) SE Sources} \\
  \midrule
  SelfEfficacy  & & \textbf{0.008 
} *** & & & & & \\
                & & \textbf{(0.000)} & & & & & \\

  MastExp       & & & \textbf{0.025} *** & & & & \textbf{0.117} *** \\
                & & & \textbf{(0.002)} & & & & \textbf{(0.006)} \\

  VicExp        & & & & \textbf{0.052} *** & & & \textbf{0.122} *** \\
                & & & & \textbf{(0.002)} & & & \textbf{(0.003)} \\

  SocPer        & & & & & \textbf{0.003} *** & & \textbf{$-$0.231} *** \\
                & & & & & \textbf{(0.002)} & & \textbf{(0.006)} \\

  PhyState      & & & & & & \textbf{0.025} *** & \textbf{0.025} *** \\
                & & & & & & \textbf{(0.002)} & \textbf{(0.004)} \\

  age           & $-$0.942 *** & $-$0.942 *** & $-$0.941 *** & $-$0.942 *** & $-$0.942 *** & $-$0.943 *** & $-$0.943 *** \\
                & (0.004) & (0.004) & (0.004) & (0.004) & (0.004) & (0.004) & (0.004) \\

  StrLen        & 0.011 *** & 0.012 *** & 0.012 *** & 0.012 *** & 0.011 *** & 0.012 *** & 0.011 *** \\
                & (0.002) & (0.002) & (0.002) & (0.002) & (0.002) & (0.002) & (0.002) \\

  DurChg        & 0.066 *** & 0.066 *** & 0.066 *** & 0.066 *** & 0.066 *** & 0.066 *** & 0.066 *** \\
                & (0.001) & (0.001) & (0.001) & (0.001) & (0.001) & (0.001) & (0.001) \\

  SocEngage     & 0.685 *** & 0.683 *** & 0.683 *** & 0.681 *** & 0.685 *** & 0.684 *** & 0.680 *** \\
                & (0.004) & (0.004) & (0.004) & (0.004) & (0.004) & (0.004) & (0.004) \\

  SocFeedback   & $-$0.059 *** & $-$0.057 *** & $-$0.057 *** & $-$0.054 *** & $-$0.058 *** & $-$0.057 *** & $-$0.052 *** \\
                & (0.002) & (0.002) & (0.002) & (0.002) & (0.002) & (0.002) & (0.002) \\

  \midrule
  Ind FE        & Yes       & Yes       & Yes       & Yes       & Yes       & Yes       & Yes \\
  Observations  & 2,567,730 & 2,567,730 & 2,567,730 & 2,567,730 & 2,567,730 & 2,567,730 & 2,567,730 \\
  N. of Users   & 175,114   & 175,114   & 175,114   & 175,114   & 175,114   & 175,114   & 175,114 \\
  Within $R^2$  & 0.336     & 0.336     & 0.336     & 0.336     & 0.336     & 0.336     & 0.337 \\
  \bottomrule
\end{tabularx}
\end{adjustwidth}
\noindent{\footnotesize{\textit{Note:} Standard errors in parentheses are clustered at the user level. Separate models capture gross associations, whereas the joint model captures conditional associations.  
*** \emph{p} < 0.001.}}

\end{table}

The separate models show positive gross associations for mastery experience ($\beta = 0.025$, $p < 0.001$), vicarious experience ($\beta = 0.052$, $p < 0.001$), social persuasion ($\beta = 0.003$, $p < 0.05$), and physiological states ($\beta = 0.025$, $p < 0.001$). The joint model reveals a more differentiated pattern. Mastery experience ($\beta = 0.117$, $p < 0.001$), vicarious experience ($\beta = 0.122$, $p < 0.001$), and physiological states ($\beta = 0.025$, $p < 0.001$) remain positively associated with exercise duration, supporting H5, H6, and H7. Social persuasion becomes negatively associated with exercise duration in the joint model ($\beta = -0.231$, $p < 0.001$), supporting H8.

{This contrast is substantively informative. Social persuasion is weakly positive when entered alone but negative after controlling for the other three sources, suggesting that its residual component may capture support-seeking, motivational vulnerability, or difficulty-related appeals for encouragement after mastery, vicarious, and physiological-affective signals are accounted for.}

{Because the joint model is the primary specification, we interpret its coefficients as approximate semi-elasticities. A one-standard-deviation increase in mastery experience, vicarious experience, physiological states, and social persuasion corresponds to about +12.4\%, +13.0\%, +2.5\%, and $-$20.6\% in exercise duration at $t+1$, or approximately +1.94, +2.03, +0.40, and $-$3.22 minutes/week using the same raw-mean reference. These modest raw-minute equivalents support criterion validity for text-based measurement but should not be interpreted as clinically meaningful changes in exercise behavior.}

Overall, these findings provide criterion-validity evidence for WEPA in multidimensional psychological measurement. They also suggest that self-efficacy expressions in online fitness communities should be interpreted as contextually embedded linguistic indicators, not as direct one-to-one replications of classical self-efficacy sources.

\subsection{{Study IV: Exploratory Relative-Week Dynamics}}
\label{subsec:study4_relative_week_dynamics}
{Study IV provides an exploratory descriptive analysis of relative-week dynamics in a naturalistic digital setting. It is not intended as a formal test of ecological validity, a hypothesis test, or the evidence of individual developmental trajectories. Instead, it examines whether WEPA-derived construct scores reveal broad and interpretable aggregate patterns when observed users are aligned by time since platform registration.}

Because users differ in activity duration and participation continuity, directly tracing individual trajectories would be misleading. Some users disengage quickly, some return after interruptions, and later relative weeks contain a more selected subset of retained users. {We therefore align users by registration week ($t = 0$), aggregate construct scores at each relative week, and interpret the results as patterns among observable users at each stage rather than as trajectories of the original registration cohort.} 

\begin{figure}[!htbp]
\centering
\includegraphics[width=\linewidth]{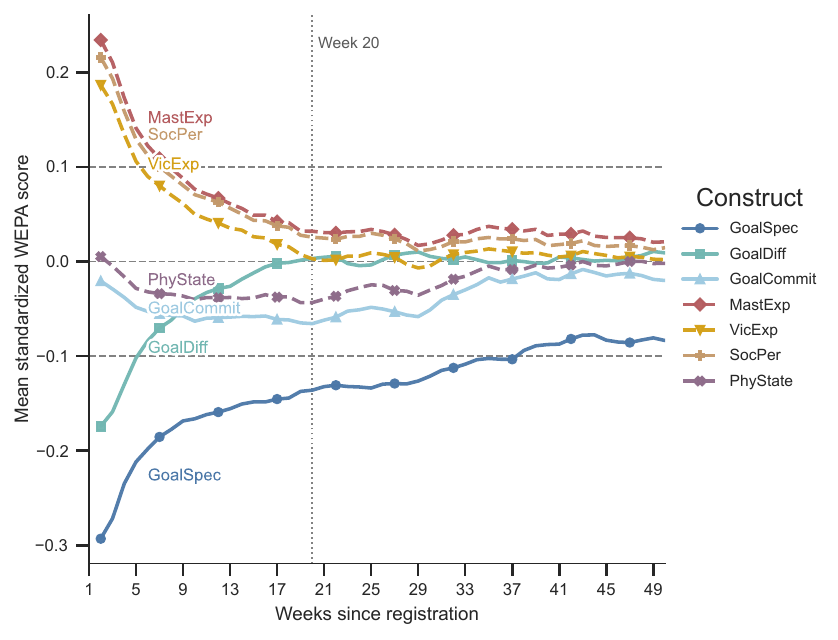}
\caption{Exploratory 
relative-week trajectories of psychological construct scores. {Each construct score is standardized using its full-sample mean and standard deviation before aggregation by relative week.}}
\label{fig:7score-trends}
\end{figure}

For each construct, this procedure produces an aggregate relative-week trajectory based on the user--week records observed during that relative week. To smooth short-term fluctuations, we apply a three-week centered moving average. {For comparability across constructs, each raw WEPA score is standardized using its full-sample mean and standard deviation before aggregation. Figure~\ref{fig:7score-trends} therefore reports mean standardized WEPA scores rather than raw projection values.}

When all seven trajectories are considered together, three descriptive patterns emerge. First, mastery experience, vicarious experience, and social persuasion show a pronounced early decline that becomes much flatter after approximately Week 20, with all three trajectories later fluctuating close to the zero line. Mastery experience generally remains the highest among the three, although their differences become smaller after the early stage.

Second, goal difficulty and goal specificity both move upward, but with different temporal profiles. Goal difficulty converges rapidly toward the zero line and remains close to it after approximately Week 20. Goal specificity starts from the lowest level and increases more gradually across the observation window. Even in later weeks, it remains below zero, making it the clearest long-term directional pattern in the figure.

Third, physiological states and goal commitment display weaker recovery-like patterns. Both decline in the early stage and later move partially toward the zero line. Compared with goal specificity and goal difficulty, these two trajectories show smaller changes and should be interpreted as mild aggregate adjustments.

{Taken together, the trajectories suggest an approximate descriptive transition around Week 20. This point is a visual reference, not a statistically estimated breakpoint. Before Week 20, several trajectories change relatively quickly; after Week 20, most constructs stabilize near the neutral range or fluctuate within a narrower band. The main long-term pattern is therefore not a uniform increase in motivational intensity, but gradual improvement in goal-related specificity alongside a broader stabilization of other construct-related expressions among retained observable users.}

\section{Robustness Checks}
\label{sec:robustness-checks}
This section reports five robustness checks covering textual sparsity, technical implementation, anchor-word perturbation, subgroup heterogeneity, and measurement-error bias. The textual-sparsity analysis examines whether the findings depend on limited valid-token information. Technical sensitivity, anchor-word perturbation, and subgroup analyses use goal commitment as a representative construct, while SIMEX covers all core constructs.

\subsection{{Sensitivity to Textual Sparsity}}
\label{subsec:sensitivity-to-textual-sparsity}

{Because WEPA scores are computed from the average vector of valid tokens, extremely short user--week texts may yield less stable construct scores. We therefore conduct a sensitivity analysis based on \textit{ValidStrLen}, the effective length of cleaned text used in WEPA scoring. First, we exclude observations with fewer than five valid tokens. Second, we compare shorter valid-text observations ($0<\textit{ValidStrLen}<20$) with longer observations ($\textit{ValidStrLen}\geq20$). Goal-setting constructs are estimated separately, consistent with Study II, and self-efficacy dimensions are estimated jointly, consistent with Study III.}

{As shown in Table~\ref{tab:textual_sparsity_sensitivity}, all focal coefficients retain their original directions and remain statistically significant after observations with fewer than five valid tokens are excluded. The shorter- and longer-text subgroup results also show stable coefficient directions. These findings suggest that the main patterns are not primarily driven by extremely sparse user--week texts.}

\begin{table}[!htb]\small
\caption{{Robustness of WEPA estimates to textual sparsity.}}
\label{tab:textual_sparsity_sensitivity}
\begin{adjustwidth}{-\extralength}{0cm}
\setlength{\cellWidtha}{\fulllength/5-2\tabcolsep+0.6in}
\setlength{\cellWidthb}{\fulllength/5-2\tabcolsep-0.4in}
\setlength{\cellWidthc}{\fulllength/5-2\tabcolsep-0.1in}
\setlength{\cellWidthd}{\fulllength/5-2\tabcolsep-0in}
\setlength{\cellWidthe}{\fulllength/5-2\tabcolsep-0.1in}

\scalebox{1}[1]{\begin{tabularx}{\fulllength}{>{\raggedright\arraybackslash}m{\cellWidtha}>{\centering\arraybackslash}m{\cellWidthb}>{\centering\arraybackslash}m{\cellWidthc}>{\centering\arraybackslash}m{\cellWidthd}>{\centering\arraybackslash}m{\cellWidthe}}
\toprule
    & \multicolumn{4}{c}{\textbf{DV:} $\boldsymbol{ExDur_{i,t+1}}$} \\
    \cmidrule{2-5}
    \textbf{Construct} & \textbf{Full Sample} & \textbf{\textit{ValidStrLen}} $\boldsymbol{\geq 5}$ & \textbf{$\boldsymbol{0<}$ \textit{ValidStrLen}} $\boldsymbol{<20}$ & \textbf{\textit{ValidStrLen}} $\boldsymbol{\geq 20}$ \\
    \midrule
    \multicolumn{5}{l}{\textit{Goal-setting constructs 
}} \\
    GoalCommit & 0.069 *** & 0.079 *** & 0.042 *** & 0.103 *** \\
     & (0.002) & (0.002) & (0.002) & (0.005) \\
    GoalSpec & 0.019 *** & 0.024 *** & 0.013 *** & 0.029 *** \\
     & (0.002) & (0.002) & (0.002) & (0.004) \\
    GoalDiff & $-$0.053 *** & $-$0.059 *** & $-$0.025 *** & $-$0.109 *** \\
     & (0.002) & (0.002) & (0.002) & (0.004) \\
    \midrule
    \multicolumn{5}{l}{\textit{Self-efficacy dimensions}} \\
    MastExp & 0.117 *** & 0.156 *** & 0.055 *** & 0.242 *** \\
     & (0.006) & (0.007) & (0.006) & (0.015) \\
    VicExp & 0.122 *** & 0.128 *** & 0.081 *** & 0.187 *** \\
     & (0.003) & (0.004) & (0.003) & (0.007) \\
    SocPer & $-$0.231 *** & $-$0.301 *** & $-$0.122 *** & $-$0.480 *** \\
     & (0.006) & (0.007) & (0.005) & (0.013) \\
    PhyState & 0.025 *** & 0.042 *** & 0.009 * & 0.068 *** \\
     & (0.004) & (0.005) & (0.004) & (0.009) \\
    \midrule
    Observations & 2,567,730 & 2,298,356 & 1,185,853 & 1,381,877 \\
    Within $R^2$ for goal-setting models & 0.336 & 0.342 & 0.260 & 0.368 \\
    Within $R^2$ for self-efficacy models & 0.337 & 0.343 & 0.261 & 0.370 \\
    Controls & Yes & Yes & Yes & Yes \\
    Individual FE & Yes & Yes & Yes & Yes \\
    \bottomrule
\end{tabularx}}
\end{adjustwidth}
\noindent{\footnotesize{\textit{Note:} Standard errors in parentheses are clustered at the user level. Goal-setting constructs are estimated separately, whereas self-efficacy dimensions are estimated jointly. All models control for DurChg, SocEngage, SocFeedback, age, and StrLen, with individual fixed effects. * \emph{p} < 0.05,  
*** \emph{p} < 0.001.}}

\end{table}

\subsection{Technical Robustness}

We test whether WEPA scores remain stable under alternative technical specifications. Using the main specification M1, defined as GloVe with 200 dimensions, a window size of 15, and the full corpus, we construct four alternatives. M2 replaces GloVe with Word2Vec, M3 reduces the vector dimension to 100, M4 reduces the context window to 10, and M5 uses a 50\% random subsample of the corpus.

For each specification, we recompute the WEPA goal commitment score and correlate it with the human benchmark. As shown in Table~\ref{tab:tech_robustness_combined}, all WEPA correlations remain above 0.86 and close to the main specification ($\rho = 0.890$), while substantially outperforming the dictionary baseline. These results indicate that WEPA's benchmark agreement is not driven by a single embedding architecture or hyperparameter setting.

\begin{table}[!htb]\footnotesize
\caption{Technical robustness across model specifications.}
\label{tab:tech_robustness_combined}
\begin{adjustwidth}{-\extralength}{0cm}
\setlength{\cellWidtha}{\fulllength/8-2\tabcolsep-0in}
\setlength{\cellWidthb}{\fulllength/8-2\tabcolsep-0in}
\setlength{\cellWidthc}{\fulllength/8-2\tabcolsep+0.05in}
\setlength{\cellWidthd}{\fulllength/8-2\tabcolsep+0.05in}
\setlength{\cellWidthe}{\fulllength/8-2\tabcolsep-0in}
\setlength{\cellWidthf}{\fulllength/8-2\tabcolsep-0in}
\setlength{\cellWidthg}{\fulllength/8-2\tabcolsep-0in}
\setlength{\cellWidthh}{\fulllength/8-2\tabcolsep-0.1in}
\scalebox{1}[1]{\begin{tabularx}{\fulllength}{>{\centering\arraybackslash}m{\cellWidtha}>{\centering\arraybackslash}m{\cellWidthb}>{\centering\arraybackslash}m{\cellWidthc}>{\centering\arraybackslash}m{\cellWidthd}>{\centering\arraybackslash}m{\cellWidthe}>{\centering\arraybackslash}m{\cellWidthf}>{\centering\arraybackslash}m{\cellWidthg}>{\centering\arraybackslash}m{\cellWidthh}}
\toprule
    \textbf{Model} & \textbf{Embedding Algorithm} & \textbf{\shortstack{Embedding\\Dimension}} & \textbf{\shortstack{Context\\Window}} & \textbf{Corpus} & \textbf{WEPA ($\rho$)} & \textbf{Dictionary ($\rho$)} & \textbf{Change vs. M1 (\%)} \\
    \midrule
    M1 (Baseline) & GloVe    & 200 & 15 & 100\% & \textbf{0.890 
} & 0.658 & ---  \\
    M2            & Word2Vec & 200 & 15 & 100\% & \textbf{0.887} & 0.658 & $-$0.3 \\
    M3            & GloVe    & 100 & 15 & 100\% & \textbf{0.881} & 0.658 & $-$1.0 \\
    M4            & GloVe    & 200 & 10 & 100\% & \textbf{0.889} & 0.658 & $-$0.1 \\
    M5            & GloVe    & 200 & 15 & 50\%  & \textbf{0.863} & 0.658 & $-$3.0 \\
    \bottomrule
\end{tabularx}}
\end{adjustwidth}
\noindent{\footnotesize{\textit{Note:} N = 518 human-annotated texts. The dictionary-based method is unaffected by model parameter settings.}}

\end{table}

\subsection{Dictionary Robustness}
\label{subsec:dictionary-robustness}

This check examines whether WEPA depends heavily on a small number of anchor words. Using goal commitment as the representative construct, we randomly remove one, two, and three positive--negative anchor-word pairs, recompute the construct scores, and re-estimate their predictive association with exercise duration at $t+1$.

As shown in Table~\ref{tab:dict_robustness}, the coefficient for goal commitment remains stable across perturbation conditions, ranging from 0.067 to 0.072 across R0--R3. The maximum deviation from the baseline estimate does not exceed 4.35\%. This pattern suggests that WEPA captures a broader semantic structure defined by the anchor-word set, not the mechanical effect of a few high-frequency words.

\begin{table}[!htb]
\caption{Robustness to anchor-word perturbation.}
\label{tab:dict_robustness}
\begin{adjustwidth}{-\extralength}{0cm}
\centering
\renewcommand{\arraystretch}{1.0}

\begin{tabularx}{\fulllength}{@{}
    >{\raggedright\arraybackslash}p{3cm}
    >{\centering\arraybackslash}X
    >{\centering\arraybackslash}X
    >{\centering\arraybackslash}X
    >{\centering\arraybackslash}X @{}}
    \toprule
    & \multicolumn{4}{c}{\textbf{DV:} $\boldsymbol{ExDur_{i,t+1}}$} \\
    \cmidrule{2-5}
    & \textbf{R0 (Baseline)} & \textbf{R1 (Remove 1 Pair)} & \textbf{R2 (Remove 2 Pairs)} & \textbf{R3 (Remove 3 Pairs)} \\
    \midrule
    GoalCommit & 0.069 *** & 0.067 *** & 0.071 *** & 0.072 *** \\
               & (0.002) & (0.002) & (0.002) & (0.002) \\
    age        & $-$0.945 *** & $-$0.945 *** & $-$0.945 *** & $-$0.945 *** \\
               & (0.004) & (0.004) & (0.004) & (0.004) \\
    StrLen     & 0.011 *** & 0.011 *** & 0.011 *** & 0.012 *** \\
               & (0.002) & (0.002) & (0.002) & (0.002) \\
    DurChg     & 0.066 *** & 0.066 *** & 0.066 *** & 0.066 *** \\
               & (0.001) & (0.001) & (0.001) & (0.001) \\
    SocEngage  & 0.681 *** & 0.681 *** & 0.681 *** & 0.680 *** \\
               & (0.004) & (0.004) & (0.004) & (0.004) \\
    SocFeedback & $-$0.054 *** & $-$0.054 *** & $-$0.053 *** & $-$0.053 *** \\
                & (0.002) & (0.002) & (0.002) & (0.002) \\
    \midrule
    Ind FE       & Yes & Yes & Yes & Yes \\
    Observations & 2,567,848 & 2,567,848 & 2,567,848 & 2,567,848 \\
    N. of Users  & 175,119 & 175,119 & 175,119 & 175,119 \\
    Within $R^2$ & 0.336 & 0.336 & 0.336 & 0.337 \\
    \bottomrule
\end{tabularx}
\end{adjustwidth}
\noindent{\footnotesize{\textit{Note:} Standard errors in parentheses and clustered at the user level.  
*** \emph{p} < 0.001.}}

\end{table}

\subsection{Cross-Group Robustness}
This check evaluates whether the association between goal commitment and future exercise duration remains stable across gender subsamples. If the association appeared only in one subgroup, WEPA's generalizability would be limited.

Table~\ref{tab:subsample_robustness} shows that goal commitment is positively associated with future exercise duration in both female and male subsamples ($\beta_{\text{female}} = 0.074$, $p < 0.001$; $\beta_{\text{male}} = 0.059$, $p < 0.001$). A between-group difference test indicates no significant coefficient difference across gender groups ($p = 0.087$). The signs of the control variables are also consistent across subsamples, supporting the stability of the focal association across gender.

\begin{table}[!htb]
\caption{Criterion validity across gender subsamples.}
\label{tab:subsample_robustness}
\begin{adjustwidth}{-\extralength}{0cm}
\centering
\renewcommand{\arraystretch}{1.0}

\begin{tabularx}{\fulllength}{@{}
    >{\raggedright\arraybackslash}p{3.2cm}
    >{\centering\arraybackslash}X
    >{\centering\arraybackslash}X
    >{\centering\arraybackslash}X @{}}
    \toprule
    & \multicolumn{3}{c}{\textbf{DV:} $\boldsymbol{ExDur_{i,t+1}}$} \\
    \cmidrule{2-4}
    & \textbf{(1) Full Sample} & \textbf{(2) Female} & \textbf{(3) Male} \\
    \midrule
    GoalCommit & 0.069 *** & 0.074 *** & 0.059 *** \\
               & (0.002) & (0.003) & (0.003) \\
    $DurChg$   & 0.066 *** & 0.071 *** & 0.057 *** \\
               & (0.001) & (0.001) & (0.002) \\
    $SocEngage$ & 0.681 *** & 0.661 *** & 0.720 *** \\
                & (0.004) & (0.005) & (0.007) \\
    $SocFeedback$ & $-$0.054 *** & $-$0.049 *** & $-$0.063 *** \\
                  & (0.002) & (0.003) & (0.004) \\
    $age$      & $-$0.945 *** & $-$0.941 *** & $-$0.949 *** \\
               & (0.004) & (0.005) & (0.006) \\
    $StrLen$   & 0.011 *** & 0.009 *** & 0.018 *** \\
               & (0.002) & (0.002) & (0.003) \\
    \midrule
    Ind FE        & Yes & Yes & Yes \\
    Observations  & 2,567,730 & 1,609,225 & 958,505 \\
    N. of Users   & 175,114 & 107,600 & 67,514 \\
    Within $R^2$  & 0.336 & 0.332 & 0.344 \\
    \bottomrule
\end{tabularx}
\end{adjustwidth}
\noindent{\footnotesize{\textit{Note:} Standard errors in parentheses and clustered at the user level.  
*** \emph{p} < 0.001.}}

\end{table}

\subsection{Estimation Robustness}
Finally, we assess whether measurement error in WEPA scores attenuates the regression estimates. Because these scores are text-based proxies for latent constructs, some measurement noise is unavoidable. Under classical measurement error, noise in explanatory variables biases coefficients toward zero.

We apply SIMEX by adding artificial measurement error with multipliers\linebreak $\lambda \in \{0, 0.5, 1, 1.5, 2\}$, running 30 simulations at each value, and extrapolating the coefficient path to $\lambda=-1$ using a quadratic function. This procedure estimates the extent to which the original coefficients are attenuated by measurement error.

Table~\ref{tab:simex_summary} shows that the naive estimates are attenuated by $-11.2\%$ to $-15.1\%$ across constructs. For goal commitment, the naive estimate is $\beta_{\text{naive}} = 0.0689$, while the SIMEX-corrected estimate is $\beta_{\text{simex}} = 0.0792$, implying a bias of $-13.0\%$. For self-efficacy, the corresponding bias is $-13.2\%$.

\begin{table}[!htb]
\caption{SIMEX estimates of measurement error.}
\label{tab:simex_summary}
\begin{adjustwidth}{-\extralength}{0cm}
\centering
\renewcommand{\arraystretch}{1.0}
\small
\begin{tabular*}{\fulllength}{@{\extracolsep{\fill}}lcccc}
    \toprule
    \textbf{Variable} & \textbf{Naive Estimate $\boldsymbol{\beta_{\text{naive}}}$} & \textbf{Corrected Estimate $\boldsymbol{\beta_{\text{simex}}}$} & \textbf{Bias (\%)} & \textbf{Interpretation} \\
    \midrule
    MastExp      & 0.0249 & 0.0286 & $-$12.9\% & Underestimated \\
    VicExp       & 0.0523 & 0.0589 & $-$11.2\% & Underestimated \\
    SocPer       & 0.0035 & 0.0041 & $-$14.6\% 
 & Underestimated \\
    PhyState     & 0.0254 & 0.0294 & $-$13.5\% & Underestimated \\
    GoalSpec     & 0.0185 & 0.0218 & $-$15.1\% & Underestimated \\
    GoalDiff    & $-$0.0526 & $-$0.0594 & $-$11.5\% & Underestimated \\
    GoalCommit   & 0.0689 & 0.0792 & $-$13.0\% & Underestimated \\
    SelfEfficacy & 0.0076 & 0.0087 & $-$13.2\% & Underestimated \\
    \bottomrule
\end{tabular*}
\end{adjustwidth}
\noindent{\scriptsize \textit{Note:} Bias (\%) $= \frac{|\beta_{\text{naive}}| - |\beta_{\text{simex}}|}{|\beta_{\text{simex}}|} \times 100\%$. Negative values indicate attenuation, meaning that naive estimates understate the absolute magnitude of the corrected estimates.}

\end{table}

\begin{figure}[!htbp]
    \begin{adjustwidth}{-\extralength}{0cm}
        \centering
        \renewcommand{\arraystretch}{0}
        \setlength{\tabcolsep}{2pt}
        \begin{tabular}{@{}cc@{}}
            \includegraphics[width=0.46\fulllength]{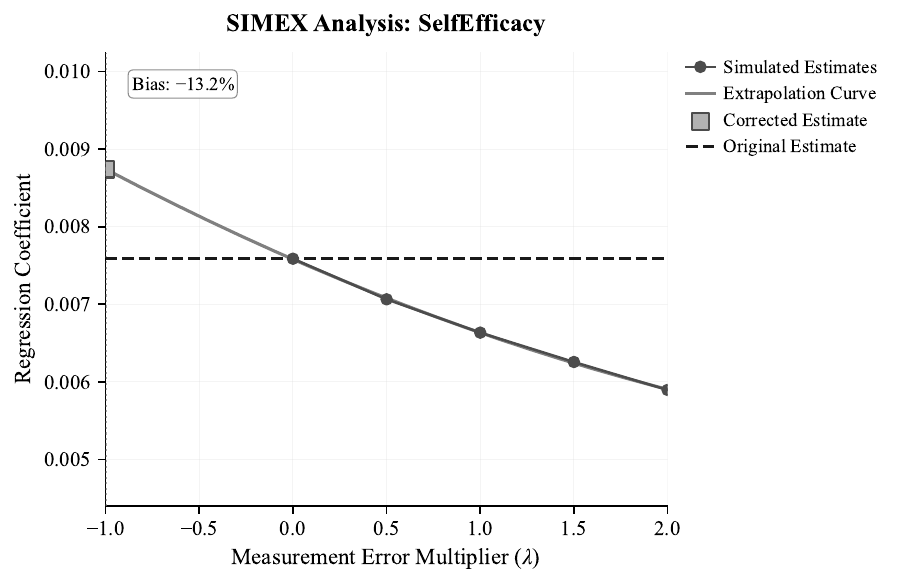} &
            \includegraphics[width=0.46\fulllength]{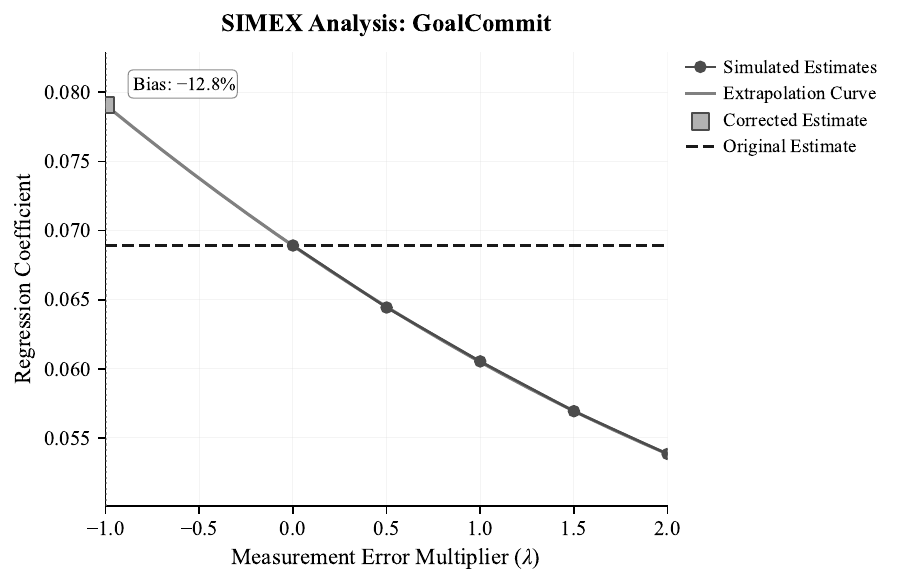} \\[6pt]
            \includegraphics[width=0.46\fulllength]{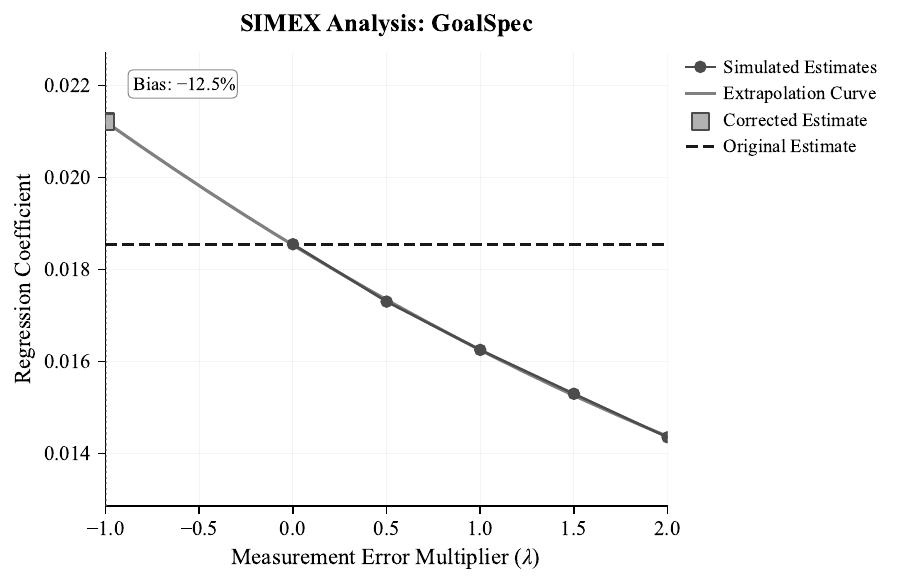} &
            \includegraphics[width=0.46\fulllength]{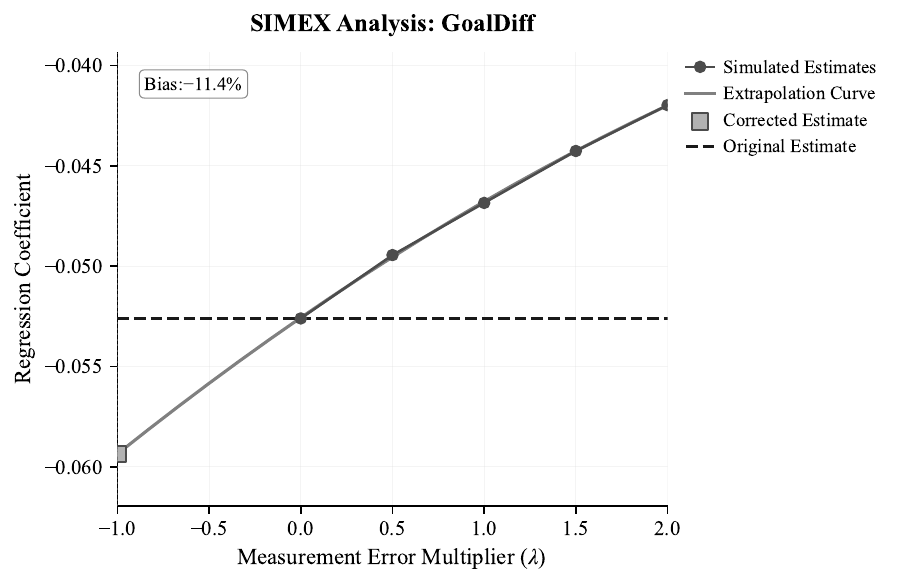} \\[6pt]
            \includegraphics[width=0.46\fulllength]{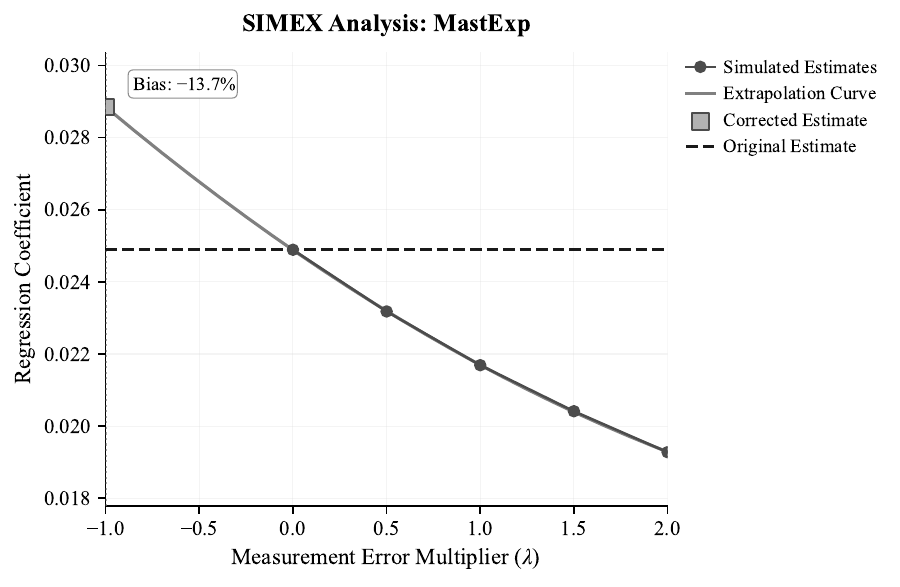} &
            \includegraphics[width=0.46\fulllength]{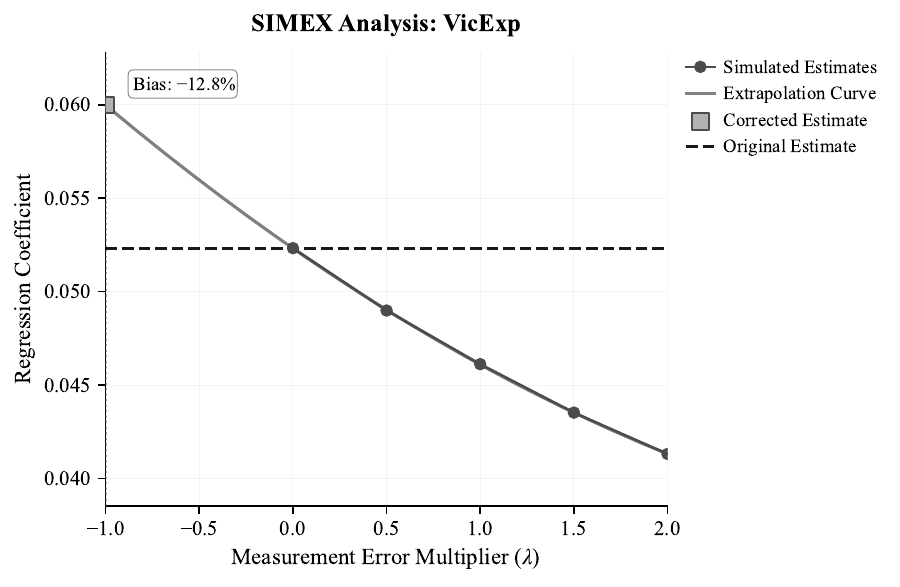} \\[6pt]
            \includegraphics[width=0.46\fulllength]{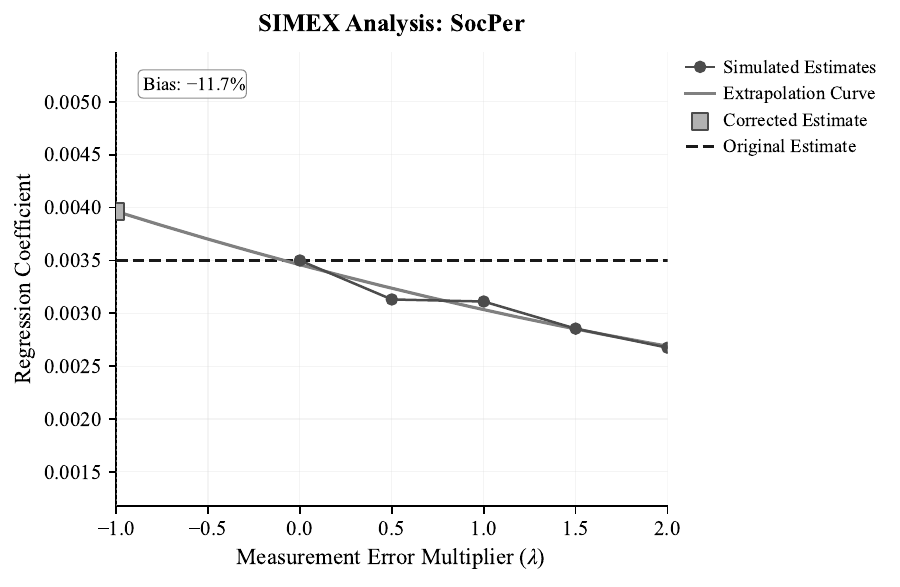} &
            \includegraphics[width=0.46\fulllength]{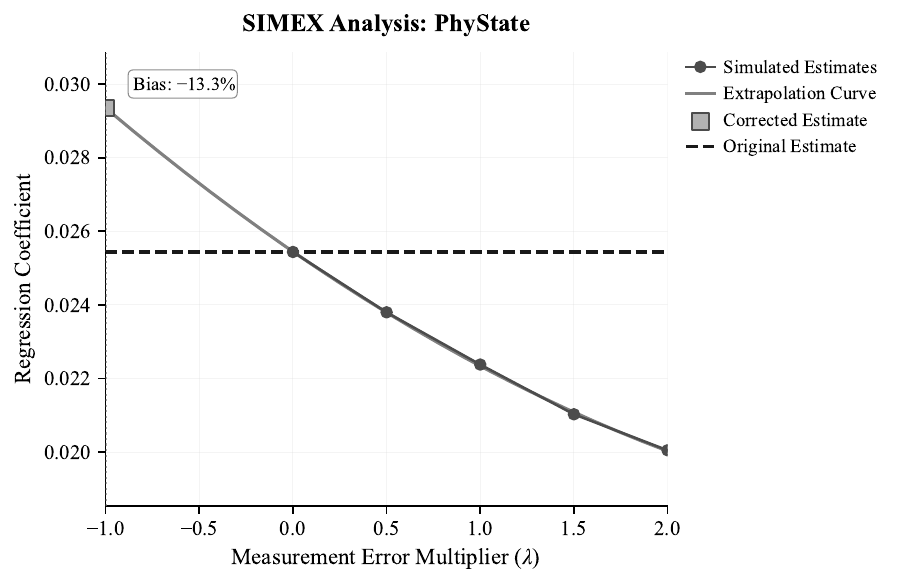} \\
        \end{tabular}
    \end{adjustwidth}
    \caption{SIMEX 
curves for core constructs.}
    \label{fig:simex_plots}
\end{figure}

Figure~\ref{fig:simex_plots} visualizes the SIMEX curves. For constructs with positive coefficients in the main models, the estimates move toward zero as measurement error increases; for constructs with negative coefficients, the estimates move toward zero from below. This pattern is consistent with classical measurement-error theory.

{SIMEX correction addresses attenuation from measurement noise, but it does not alter the observational nature of the study or establish causality.} Overall, the corrected estimates suggest that the baseline coefficients are conservative. Measurement noise modestly attenuates association strength without changing coefficient signs, supporting the robustness of WEPA-based construct scores.

\section{Discussion}
\subsection{Main Findings}
This study developed the Word Embedding Projection Approach (WEPA) and evaluated its usefulness for theory-driven psychological measurement from social media text. Using long-panel data from the Keep fitness community, the results provide convergent evidence on measurement performance, behavioral relevance, temporal comparability, and exploratory longitudinal applicability. WEPA showed strong agreement with human annotation, achieved broader text coverage than the dictionary-based baseline, and produced construct scores meaningfully associated with subsequent exercise behavior. The relative-week analysis further showed that WEPA can recover interpretable aggregate temporal patterns in user-generated text. These findings support the usefulness of WEPA as an interpretable measurement approach. At the same time, they should be understood as evidence of measurement performance, predictive associations, and exploratory descriptive utility, not as evidence of causal effects or strict measurement invariance.

\textls[-15]{Several substantive patterns emerged across constructs. For goal-setting constructs, goal commitment and goal specificity were positively associated with later exercise duration, consistent with Goal-Setting Theory \citep{Locke2019The}. Goal difficulty showed a negative association, reflecting the platform-specific meaning of difficulty-related expressions, which often capture perceived obstacles, fatigue, or reduced persistence. For self-efficacy, mastery experience, vicarious experience, and physiological states were positively associated with later exercise behavior, while social persuasion showed a negative association. {This result does not imply that social persuasion weakens self-efficacy. It suggests that users' own mentions of encouragement or support may become salient when they are struggling, seeking reassurance, or experiencing motivational vulnerability \citep{Bandura1977SelfEfficacy}.}}

{The temporal analysis adds a descriptive perspective on aggregate psychological dynamics. Construct-related language changed systematically across participation stages, with early declines followed by stabilization for several dimensions and gradual increases in goal specificity and goal difficulty. These patterns do not establish individual developmental trajectories, because later relative weeks contain a more selected group of continuing users and participation intensity varies across users. They nevertheless illustrate how WEPA can describe aggregate changes in theory-relevant psychological language in naturalistic digital settings.}

\subsection{Contributions to Theory and Measurement}
This study makes three contributions to computational behavioral science and psychological measurement. First, it introduces WEPA as a theory-driven and interpretable approach for measuring psychological constructs from social media text. WEPA operationalizes constructs as semantic projections along theoretically grounded bipolar axes, making it most suitable for constructs that can be meaningfully represented through contrastive semantic poles. {By linking anchor words, semantic axes, and construct scores, WEPA connects psychological theory with distributional semantics and offers an interpretable complement to dictionary-based methods and deep learning models \citep{kozlowski2019GeometryCulture, grand2022SemanticProjection}. Its application to more than 2.6 million user--week observations demonstrates practical feasibility for large-scale longitudinal text analysis, while broader scalability across platforms, languages, and computational environments remains a direction for future evaluation.}

Second, the study contributes to validation practice by treating text-based psychological measurement as a layered validation problem. The validation strategy combines benchmark-based evaluation, predictive associations with subsequent behavior, temporal-stability diagnostics, and exploratory longitudinal analysis. These components address different questions, including whether WEPA aligns with human judgment, whether WEPA-derived scores are associated with later behavior, whether semantic axes remain sufficiently comparable across time, and whether construct-related language shows interpretable aggregate temporal patterns. This structure helps to avoid treating annotation agreement, behavioral association, temporal comparability, and descriptive dynamics as one undifferentiated form of validity \citep{Flake2017ConstructValidation, Kyle2020Tools}.  

{Third, the study clarifies how classical psychological constructs should be interpreted when measured through user-generated text. The findings broadly support the relevance of Goal-Setting Theory and Social Cognitive Theory in digital fitness communities, but they also show that text-based scores capture expressed psychological salience rather than the external motivational conditions described in classical theories. This distinction is especially important for goal difficulty and social persuasion. Difficulty-related language may reflect perceived obstacles or fatigue, while persuasion-related language may reflect users' own mentions of encouragement or support. WEPA therefore provides a structured way to study theory-relevant psychological language while keeping the interpretation of digital trace measures within appropriate boundaries.}

\subsection{Practical Implications}

{WEPA has practical implications for monitoring psychological signals in digital health platforms. By converting user-generated text into interpretable construct scores, it can help to identify aggregate signals that are difficult to observe through behavioral logs alone. Changes in expressed goal commitment, perceived difficulty, or self-efficacy-related language may indicate motivational strain, declining persistence, or emerging support needs. These signals can support the platform-level monitoring and retrospective evaluation of community dynamics.}

WEPA may also support more theory-informed intervention design. Platforms could move beyond generic engagement metrics and examine whether users' language reflects clearer goals, stronger commitment, positive efficacy-related experiences, or greater expressed difficulty. Such information may inform feedback messages, goal-setting prompts, community support features, personalized encouragement strategies, and the timing of intervention delivery.

These applications require caution. WEPA scores reflect expressed psychological salience in text, not direct diagnoses of users' internal states or objective motivational conditions. This approach is most appropriate for aggregate analysis, content design, and platform-level decision support. Individual-level applications require additional validation, ethical safeguards, and transparency to avoid overinterpretation or intrusive \mbox{behavioral profiling}.

\subsection{Limitations and Future Research}
\label{subsec:limitations-future-research}
{Several limitations should be acknowledged. First, the generalizability of WEPA requires further validation. The present study is based on a single Chinese fitness-oriented social media platform, and both the embedding space and anchor-word dictionaries were developed for this specific linguistic and behavioral context. This localization strengthens construct validity in the Keep community but limits direct portability across platforms, languages, and cultural settings. In addition, the network-based sampling procedure and the exclusion of users with insufficient observation records may overrepresent socially active and more persistent users. Future research should examine whether the WEPA workflow can be replicated in other digital environments and how anchor-word dictionaries can be adapted through cross-domain validation, cross-language alignment, theory-guided seed words, large-language-model expansion, and expert review.}

{Second, the measurement evidence remains bounded in scope. Study I provides anchor-word validation, human-coded benchmarks for selected constructs, and cross-temporal semantic-axis stability checks. These analyses support longitudinal comparability, although they do not provide a full test of measurement invariance in the strict psychometric sense \citep{Liu2017Testing, Leitgob2022Measurement}. WEPA is also most suitable for constructs that can be meaningfully represented through contrastive semantic poles; extending the approach to non-bipolar constructs remains a direction for future research. Some construct dimensions, especially the four self-efficacy sources, show high empirical correlations, suggesting substantial shared variance across theoretically distinct dimensions in platform text. Although the textual-sparsity sensitivity analysis suggests that the main results are not driven by extremely sparse texts, weekly aggregation still involves a trade-off between measurement stability and temporal precision. Future work should develop larger annotated benchmarks, examine discriminant validity across related constructs, and explore alternative aggregation strategies.}

Third, future applications of WEPA need to improve the scalability of anchor-word development and expand beyond text-only measurement. In the current study, anchor-word dictionaries were developed through theory-guided selection and expert validation, which supports construct validity but remains labor-intensive when WEPA is extended to new constructs, platforms, or languages. Large language models may help to generate candidate anchor words from theory-guided seed terms, explain semantic distinctions among candidates, and support cross-language adaptation, although expert review and empirical validation remain necessary. Multimodal large language models also make it increasingly feasible to align text, images, audio, video, and behavioral traces within a shared semantic space. Future research could extend WEPA from text-based measurement to multimodal semantic projection and move psychological measurement from weekly aggregation toward daily or event-level analysis \citep{Sudarsanam2025Representation, Zhu2024A}. {Because the present study is observational, the reported associations should be interpreted as predictive associations. Future work could combine WEPA with experimental or quasi-experimental designs to support stronger causal inference.}

{Overall, WEPA should be understood as a promising and context-sensitive measurement workflow, not as a fully general solution for psychological measurement from all social media text.} The present study provides evidence from Chinese fitness-oriented social media text. Broader claims require replication across platforms, languages, populations, and theoretical constructs.
\vspace{6pt}

\section*{Author Contributions}
Conceptualization, X.D.; methodology, X.D.; software, X.D.; validation, X.D.; formal analysis, X.D.; investigation, X.D.; data curation, X.D.; writing---original draft preparation, X.D.; writing---review and editing, X.D.; supervision, Y.L.; project administration, Y.L.; funding acquisition, Y.L. All authors have read and agreed to the published version of the manuscript.

\section*{Funding}
This research received no external funding.

\section*{Institutional Review}
This study is fully exempt from institutional ethical review under the Measures for the 
 Ethical Review of Life Science and Medical Research Involving Human Beings (Order No. 4 [2023], National Health Commission of the People’s Republic of China). According to these Measures, research that uses publicly available data and involves no intervention, no interaction with human participants, and no collection of identifiable personal information falls within the statutory exemption scope. The data used in this study consist solely of publicly accessible social media text that was voluntarily posted by users. Prior to analysis, all data underwent complete anonymization and de‑identification, with the permanent removal of all personal identifiers (including usernames, IDs, contact information, and any traceable attributes). No human subjects, experiments, surveys, or interventions were conducted, and all analyses were performed exclusively at the aggregated group level. Therefore, no individual can be identified directly or indirectly from the processed dataset. In compliance with the above national regulatory framework, ethical approval was not required for this study.

\section*{Informed Consent}
Ethical approval for the waiver of informed consent was granted because the study employed only publicly accessible and de-identified secondary data, did not involve direct contact with users, and did not report any identifiable individual-level information.

\section*{Data Availability}
The code for WEPA scoring and the complete Chinese anchor-word dictionaries are available in the open-source Python 
package \texttt{cntext} (version 2.2.0): \url{https://github.com/hiDaDeng/cntext}, accessed on 
12 May 2026. The repository includes installation instructions, a minimal WEPA example, API descriptions, and basic tests for semantic-axis construction and text scoring. The raw platform data contain user-generated content and are not publicly redistributed to protect user privacy.

\section*{Acknowledgments}
The authors are grateful to the anonymous reviewers for their valuable comments and constructive suggestions, which have greatly improved the quality of this manuscript.

\section*{Conflicts of Interest}
The authors declare no conflicts of interest.

\bibliographystyle{apacite}
\bibliography{references}

@article{Kyle2020Tools,
title={Tools for assessing the psychometric adequacy of latent variables in conservation research},
author={G. Kyle and Adam C. Landon and J. Vaske and K. Wallen},
journal={Conservation Biology},
year={2020},
volume={34},
number={6},
pages={1353--1363},
doi={10.1111/cobi.13625}
}

@article{Kadhim2019Survey,
title={Survey on supervised machine learning techniques for automatic text classification},
author={Ammar Ismael Kadhim},
journal={Artificial Intelligence Review},
year={2019},
volume={52},
pages={273-292},
doi={10.1007/s10462-018-09677-1}
}

@article{Speer2024Taking,
title={Taking It Easy: Off-the-Shelf Versus Fine-Tuned Supervised Modeling of Performance Appraisal Text},
author={Andrew B. Speer and James Perrotta and Tobias L. Kordsmeyer},
journal={Organizational Research Methods},
year={2024},
volume={29},
pages={73 - 91},
doi={10.1177/10944281241271249}
}

@article{Zhou2024Larger,
title={Larger and more instructable language models become less reliable},
author={Lexin Zhou and Wout Schellaert and Fernando Martínez-Plumed and Yael Moros-Daval and César Ferri and J. Hernández-Orallo},
journal={Nature},
year={2024},
volume={634},
pages={61 - 68},
doi={10.1038/s41586-024-07930-y}
}

@article{Sclar2023Quantifying,
title={Quantifying Language Models' Sensitivity to Spurious Features in Prompt Design or: How I learned to start worrying about prompt formatting},
author={Melanie Sclar and Yejin Choi and Yulia Tsvetkov and Alane Suhr},
journal={ArXiv},
year={2023},
volume={abs/2310.11324},
doi={10.48550/arxiv.2310.11324}
}

@article{Demszky2023Using,
title={Using large language models in psychology},
author={Dorottya Demszky and Diyi Yang and David S. Yeager and Christopher J. Bryan and Margarett Clapper and Susannah Chandhok and J. Eichstaedt and Cameron A. Hecht and Jeremy P. Jamieson and Meghann Johnson and Michaela Jones and Danielle Krettek-Cobb and Leslie Lai and Nirel JonesMitchell and Desmond C. Ong and C. Dweck and James J. Gross and James W. Pennebaker},
journal={Nature Reviews Psychology},
year={2023},
volume={2},
pages={688--701},
doi={10.1038/s44159-023-00241-5}
}

@article{Eichstaedt2020ClosedOpen,
title={Closed- and open-vocabulary approaches to text analysis: a review, quantitative comparison, and recommendations.},
author={J. C. Eichstaedt and Margaret L. Kern and D. B. Yaden and H. A. Schwartz and S. Giorgi and G. Park and C. A. Hagan and V. A. Tobolsky and Laura K. Smith and A. Buffone and J. Iwry},
journal={Psychological Methods},
year={2020},
volume={26},
number={4},
pages={398--427},
doi={10.1037/met0000349}
}

@article{Eichstaedt2020Tracking,
title={Tracking fluctuations in psychological states using social media language: A case study of weekly emotion},
author={J. Eichstaedt and A. Weidman},
journal={European Journal of Personality},
year={2020},
volume={34},
number={5},
pages={845--858},
doi={10.1002/per.2261}
}

@article{Flake2017ConstructValidation,
title={Construct validation in social and personality research},
author={J. Flake and J. Pek and Eric Hehman},
journal={Social Psychological and Personality Science},
year={2017},
volume={8},
number={4},
pages={370--378},
doi={10.1177/1948550617693063}
}

@article{Hartmann2019Comparing,
title={Comparing automated text classification methods},
author={Jochen Hartmann and Juliana Huppertz and Christina Schamp and Mark Heitmann},
journal={International Journal of Research in Marketing},
year={2019},
volume={36},
number={1},
pages={20--38},
doi={10.1016/j.ijresmar.2018.09.009}
}

@article{Humphreys2017Automated,
	author   = {Humphreys, Ashlee and Wang, Rebecca Jen-Hui},
	title    = {Automated text analysis for consumer research},
	journal  = {Journal of Consumer Research},
	volume   = {44},
	number   = {6},
	pages    = {1274--1306},
	year     = {2017},
	month    = {09},
	issn     = {0093-5301},
	doi      = {10.1093/jcr/ucx104},
	url      = {https://doi.org/10.1093/jcr/ucx104},
	eprint   = {https://academic.oup.com/jcr/article-pdf/44/6/1274/29010413/ucx104.pdf}
}

@article{Kalnins2023The,
title={The {VIF} score. What is it good for? Absolutely nothing},
author={Arturs Kalnins and Kendall Praitis Hill},
journal={Organizational Research Methods},
year={2023},
volume={28},
pages={58--75},
doi={10.1177/10944281231216381}
}

@article{Kjell2019Semantic,
title={Semantic measures: Using natural language processing to measure, differentiate, and describe psychological constructs},
author={O. Kjell and Katarina Kjell and Danilo Garcia and S. Sikström},
journal={Psychological Methods},
year={2019},
volume={24},
number={1},
pages={92--115},
doi={10.1037/met0000191}
}

@misc{Sudarsanam2025Representation,
  title={Representation Learning for Semantic Alignment of Language, Audio, and Visual Modalities}, 
  author={Sudarsanam, Parthasaarathy and Mart{\'i}n-Morat{\'o}, Irene and Virtanen, Tuomas},
  year={2025},
  eprint={2505.14562},
  archivePrefix={arXiv},
  primaryClass={cs.SD},
  url={https://arxiv.org/abs/2505.14562},
  howpublished={arXiv preprint},
  note={arXiv:2505.14562. Available at: https://arxiv.org/abs/2505.14562}
}

@inproceedings{Yao2017Dynamic,
author = {Yao, Zijun and Sun, Yifan and Ding, Weicong and Rao, Nikhil and Xiong, Hui},
title = {Dynamic Word Embeddings for Evolving Semantic Discovery},
year = {2018},
isbn = {9781450355810},
publisher = {Association for Computing Machinery},
address = {New York, NY, USA},
url = {https://doi.org/10.1145/3159652.3159703},
doi = {10.1145/3159652.3159703},
booktitle = {Proceedings of the Eleventh ACM International Conference on Web Search and Data Mining},
pages = {673–681},
numpages = {9},
location = {Marina Del Rey, CA, USA},
series = {WSDM '18}
}

@article{Zhu2024A,
title={A review of key technologies for emotion analysis using multimodal information},
author={Xianxun Zhu and Chaopeng Guo and Heyang Feng and Yao Huang and Yichen Feng and Xiangyang Wang and Rui Wang},
journal={Cognitive Computation},
year={2024},
volume={16},
pages={1504--1530},
doi={10.1007/s12559-024-10287-z}
}

@article{golder2014DigitalFootprints,
  title={Digital footprints: Opportunities and challenges for online social research},
  author={Golder, Scott A and Macy, Michael W},
  journal={Annual review of sociology},
  volume={40},
  number={1},
  pages={129--152},
  year={2014},
  publisher={Annual Reviews},
  doi={10.1146/annurev-soc-071913-043145}
}

@article{grand2022SemanticProjection,
	title     = {Semantic projection recovers rich human knowledge of multiple object features from word embeddings},
	author    = {Grand, Gabriel and Blank, Idan Asher and Pereira, Francisco and Fedorenko, Evelina},
	journal   = {Nature Human Behaviour},
	pages     = {975--987},
	year      = {2022},
	volume={6},
	publisher = {Nature Publishing Group},
	doi={10.1038/s41562-022-01316-8}
}

@article{herderich2024computational,
  title={A computational method to reveal psychological constructs from text data},
  author={Herderich, Alina and Freudenthaler, Heribert H. and Garcia, David},
  journal={Psychological Methods},
  year={2024},
  doi={10.1037/met0000700}
}

@misc{deng2022cntext,
  author       = {Deng, X. and Nan, P.},
  title        = {cntext: A {Python} tool for text mining},
  year         = {2022},
  publisher    = {Zenodo},
  doi          = {10.5281/zenodo.7063523},
  url          = {https://github.com/hiDaDeng/cntext},
  howpublished = {Computer software},
  note         = {Zenodo. Available at: https://github.com/hiDaDeng/cntext (accessed 12 May 2026)}
}

@article{berger2020UnitingTribes,
	title     = {Uniting the tribes: Using text for marketing insight},
	author    = {Berger, Jonah and Humphreys, Ashlee and Ludwig, Stephan and Moe, Wendy W and Netzer, Oded and Schweidel, David A},
	journal   = {Journal of Marketing},
	volume    = {84},
	number    = {1},
	pages     = {1--25},
	year      = {2020},
	doi       = {10.1177/0022242919873106},
	publisher = {Sage Publications Sage CA: Los Angeles, CA}
}

@article{kozlowski2019GeometryCulture,
	title     = {The geometry of culture: Analyzing the meanings of class through word embeddings},
	author    = {Kozlowski, Austin C and Taddy, Matt and Evans, James A},
	journal   = {American Sociological Review},
	volume    = {84},
	number    = {5},
	pages     = {905--949},
	year      = {2019},
	doi       = {10.1177/0003122419877135},
	publisher = {SAGE Publications Sage CA: Los Angeles, CA}
}

@article{nikadon2025BERTAgent-MeasurePsyMetrics,
  title={{BERTAgent}: The development of a novel tool to quantify agency in textual data.},
  author={Nikadon, Jan and Suitner, Caterina and Erseghe, Tomaso and D{\v{z}}anko, Lejla and Formanowicz, Magdalena},
  journal={Journal of Experimental Psychology: General},
  year={2025},
  volume={154},
  number={7},
  pages={1855--1877},
  doi={10.1037/xge0001740},    
  publisher={American Psychological Association},
}

@misc{pennebaker2015development,
  author       = {Pennebaker, James W. and Boyd, Ryan L. and Jordan, Kayla and Blackburn, Kate},
  title        = {The development and psychometric properties of {LIWC2015}},
  year         = {2015},
  publisher    = {University of Texas at Austin},
  doi          = {10.15781/T29G6Z},
  url          = {https://repositories.lib.utexas.edu/handle/2152/31333},
  howpublished = {Technical Report},
  note         = {University of Texas at Austin. Available at: https://repositories.lib.utexas.edu/handle/2152/31333 (accessed 12 May 2026)}
}

@article{aceves2024mobilizing,
	title     = {Mobilizing conceptual spaces: How word embedding models can inform measurement and theory within organization science},
	author    = {Aceves, Pedro and Evans, James A},
	journal   = {Organization Science},
	volume    = {35},
	number    = {3},
	pages     = {788--814},
	year      = {2024},
	doi       ={10.1287/orsc.2023.1686},
	publisher = {INFORMS}
}

@article{Douven2011Vagueness,
title={Vagueness: A Conceptual Spaces Approach},
author={I. Douven and L. Decock and R. Dietz and P. Égré},
journal={Journal of Philosophical Logic},
year={2011},
volume={42},
pages={137--160},
doi={10.1007/s10992-011-9216-0}
}

@article{Barchard2024Testing,
title={Testing Bipolarity},
author={K. Barchard and James M. Carroll and Shawn Reynolds and James A. Russell},
journal={Psychological Methods},
year={2024},
doi={10.1037/met0000707}
}

@article{Reise2013Scoring,
title={Scoring and Modeling Psychological Measures in the Presence of Multidimensionality},
author={S. Reise and Wes Bonifay and M. G. Haviland},
journal={Journal of Personality Assessment},
year={2013},
volume={95},
pages={129--140},
doi={10.1080/00223891.2012.725437}
}

@article{Rodriguez2016Applying,
title={Applying Bifactor Statistical Indices in the Evaluation of Psychological Measures},
author={Anthony Rodriguez and S. Reise and M. G. Haviland},
journal={Journal of Personality Assessment},
year={2016},
volume={98},
pages={223--237},
doi={10.1080/00223891.2015.1089249}
}

@article{Bandura1977SelfEfficacy,
title={Self-efficacy: Toward a unifying theory of behavioral change.},
author={A. Bandura},
journal={Psychological Review},
year={1977},
volume={84},
number={2},
pages={191--215},
doi={10.1037/0033-295x.84.2.191}
}

@article{O’Brien2007A,
title={A caution regarding rules of thumb for variance inflation factors},
author={Robert M. O’Brien},
journal={Quality \& Quantity},
year={2007},
volume={41},
pages={673--690},
doi={10.1007/s11135-006-9018-6}
}

@article{Locke2019The,
title={The development of goal setting theory: A half century retrospective.},
author={E. A. Locke and G. Latham},
journal={Motivation Science},
year={2019},
volume={5},
number={2},
pages={93--105},
doi={10.1037/mot0000127}
}

@article{Brandstatter2021Persistence,
title={Persistence and Disengagement in Personal Goal Pursuit.},
author={V. Brandstätter and K. Bernecker},
journal={Annual Review of Psychology},
year={2022},
volume={73},
pages={271--299},
doi={10.1146/annurev-psych-020821-110710}
}

@article{Cassani2021Words,
  title   = {Words with Consistent Diachronic Usage Patterns are Learned Earlier: A Computational Analysis Using Temporally Aligned Word Embeddings},
  author  = {Cassani, Giovanni and Bianchi, Federico and Marelli, Marco},
  journal = {Cognitive Science},
  year    = {2021},
  volume  = {45},
  number  = {4},
  pages   = {e12963},
  doi     = {10.1111/cogs.12963}
}

@article{Wirth2009The,
title={The impact of goal specificity and goal type on learning outcome and cognitive load},
author={Joachim Wirth and Josef Künsting and D. Leutner},
journal={Computer in Human Behavior},
year={2009},
volume={25},
number={2},
pages={299--305},
doi={10.1016/j.chb.2008.12.004}
}

@article{Senko2005Achievement,
title={Achievement goals, task performance, and interest: Why perceived goal difficulty matters},
author={Corwin Senko and J. Harackiewicz},
journal={Personality and Social Psychology Bulletin},
year={2005},
volume={31},
number={12},
pages={1739--1753},
doi={10.1177/0146167205281128}
}

@article{Liu2017Testing,
title={Testing measurement invariance in longitudinal data with ordered-categorical measures},
author={Yu Liu and R. Millsap and S. West and J. Tein and R. Tanaka and Kevin J. Grimm},
journal={Psychological Methods},
year={2017},
volume={22},
pages={486–-506},
doi={10.1037/met0000075}
}

@article{Leitgob2022Measurement,
title={Measurement invariance in the social sciences: Historical development, methodological challenges, state of the art, and future perspectives.},
author={Leitg{\"o}b, H. and Seddig, Daniel and Asparouhov, T. and Behr, Doroth{\'e}e and Davidov, E. and De Roover, Kim and Jak, S. and Meitinger, Katharina and Menold, Natalja and Muth{\'e}n, B. and Rudnev, M. and Schmidt, Peter and {van de Schoot}, R.},
journal={Social Science Research},
year={2022},
volume={110},
pages={35--65},
doi={10.1016/j.ssresearch.2022.102805}
}

@article{Lee2023LDA,
author = {Lee, Jea Woog and Han, Doug Hyun},
title = {Data Analysis of Psychological Approaches to Soccer Research: Using LDA Topic Modeling},
journal = {Behavioral Sciences},
volume = {13},
number = {10},
article-number = {787},
pages = {787},
year = {2023},
doi = {10.3390/bs13100787}
}

@article{Luisa2024PsyChange,
author = {Orrù, Luisa and Cuccarini, Marco and Moro, Christian and Turchi, Gian Piero},
title = {Clinical Questions and Psychological Change: How Can Artificial Intelligence Support Mental Health Practitioners?},
journal = {Behavioral Sciences},
volume = {14},
year = {2024},
number = {12},
article-number = {1225},
pages = {1225},
doi = {10.3390/bs14121225}
}

@article{Feng2023SNS,
author = {Feng, Ran and Ivanov, Alex},
title = {Gender Differences in Emotional Valence and Social Media Content Engagement Behaviors in Pandemic Diaries: An Analysis Based on Microblog Texts},
journal = {Behavioral Sciences},
volume = {13},
year = {2023},
number = {1},
article-number = {34},
pages = {34},
doi = {10.3390/bs13010034}
}

@article{Zhang2024PsyOwnershipUGC,
author = {Zhang, Nan and Hu, Wen},
title = {Do Psychological Ownership and Communicative Presence Matter? Examining How User-Generated Content in E-Commerce Live Streaming Influences Consumers’ Purchase Intention},
journal = {Behavioral Sciences},
volume = {14},
year = {2024},
number = {8},
article-number = {696},
pages = {696},
doi = {10.3390/bs14080696}
}

@article{sultan2023leaving,
  title={Leaving traces behind: Using social media digital trace data to study adolescent wellbeing},
  author={Sultan, Mubashir and Scholz, Christin and van den Bos, Wouter},
  journal={Computers in Human Behavior Reports},
  volume={10},
  pages={100281},
  year={2023},
  publisher={Elsevier},
  doi = {10.1016/j.chbr.2023.100281},
}

\end{document}